# Nanoscale friction of manganite superlattice films controlled by layer thickness and fluorine content

Niklas A. Weber[a], Miru Lee[b], Florian Schönewald[a], Leonard Schüler[c], Vasily Moshnyaga[c], Matthias Krüger[b], and Cynthia A. Volkert[a,d,e,*]

[a] Institute of Materials Physics, University of Göttingen, Friedrich-Hund-Platz 1, Göttingen 37077, Germany

[b] Institute for Theoretical Physics, University of Göttingen, Friedrich-Hund-Platz 1, 37077 Göttingen, Germany

[c] 1st Physics Institute, University of Göttingen, Friedrich-Hund-Platz 1, 37077 Göttingen, Germany

[d] The International Center for Advanced Studies of Energy Conversion (ICASEC), University of Göttingen, Göttingen 37077, Germany

[e] International Institute for Carbon Neutral Energy Research (WPI-I2CNER), Kyushu University, 744 Motooka, Nishi-ku, Fukuoka 819-0935, Japan

* Email: cynthia.volkert@phys.uni-goettingen.de

Author Information:

Niklas A. Weber (niklas.weber@mailbox.org)

Miru Lee (zeros0121@gmail.com),

Florian Schönewald (florian.schoenewald@icloud.com)

Leonard Schüler (mail.leonard.schueler@web.de)

Vasily Moshnyaga (vmosnea@gwdg.de)

Matthias Krüger (matthias.kruger@uni-goettingen.de)

Cynthia A. Volkert (cynthia.volkert@phys.uni-goettingen.de)

Abstract

Frictional losses play a major role in global energy consumption, making control of friction across length scales essential for improving efficiency. Since friction at the mesoscale often originates from processes at the nanoscale, understanding and controlling nanoscale friction is key to designing low-friction, high-performance materials. Here, we investigate nanoscale friction in $[LaMnO_3]_m/[SrMnO_3]_n$ superlattice films using lateral force microscopy, focusing on the effects of fluorine doping and top-layer thickness. Across all films, the friction force scales linearly with the sum of the applied normal load and adhesion force. While absolute friction forces vary locally due to adhesion variations, the friction coefficient remains consistent for each film but is systematically influenced by fluorine concentration and top-layer thickness. The thickness dependence indicates that frictional energy dissipation extends up to ~5 nm below the surface, underscoring the role of subsurface structure. We attribute this dissipation to viscoelastic losses in the stress field and evanescent waves generated by the sliding tip, which quantitatively explain the observed friction coefficients. These results demonstrate that, once adhesion is accounted for, the friction coefficient is a reproducible material property that can be systematically tuned—offering a general strategy for tailoring frictional behavior through controlled surface and subsurface modifications.

## 1. Introduction

Controlling friction at a sliding contact has been a long-standing challenge in technology and society. A large fraction of the energy that society consumes is lost in frictional dissipation,[1] but at the same time, friction is essential to traction and mechanical control. Ideally, ways will be found to turn friction off when it is not needed and to turn it on when forces need to be imparted on the object. A significant amount of research has focused on trying to control friction, and with the emergence of atomic force microscopy (AFM) in the 1980s,[2] friction at a nanoscale contact could be studied, allowing insights into fundamental mechanisms of energy loss.

A general framework for understanding dry, low-wear friction has emerged out of these studies. Most importantly, the ground-breaking work of Bowden and Tabor showed that a mesoscopic contact consists of many individual asperity contacts. They argued that the measured friction force is proportional to the *true* area of contact and not the apparent contact area.[3] Numerous experiments have tested this idea by measuring both the contact area and friction force. While significant scatter is common in the data, many studies support the proportionality between friction force and true contact area.

Consequently, discussions of friction between hard materials have often focused on dissipation mechanisms occurring directly in the contact interface.[4] However, recent evidence points to potential *subsurface* contributions to friction in hard materials. These experimental observations are typically based on changes in friction resulting from modifications of material properties via temperature variation,[5–8] contact pressure,[9] or bias voltage/carrier density.[10–13] In several cases, significant changes in friction correlate with changes in phononic and electronic degrees of freedom in the material, and a variety of models have been proposed to attempt to account for these. However, in most cases, the predicted dissipation from phononic and electronic mechanisms is often too small to match the experimental results.[14] Moreover, most studies are complicated by the fact that the applied fields simultaneously alter both the surface and subsurface properties of the specimen, making it difficult to distinguish whether the observed friction changes originate from subsurface energy dissipation or from modified interfacial dynamics at the sliding contact.

In addition to phononic and electronic degrees of freedom, viscoelastic contributions from sub-surface regions to friction have also been considered, particularly for the case of elastomers such as rubber. As early examples, internal friction or elastic hysteresis losses have been proposed as the main origin of rolling friction on rubber and several metals[15] and of sliding friction on rubber.[16] Various quantitative models have been proposed for rubber friction and critically compared,[17] that are often based on the idea that friction forces and coefficients are the sum of two terms: a term proportional to the true contact area that scales with adhesion hysteresis and a term proportional to the stressed volume and the internal friction of the material.

An experimental approach that allows volume and surface effects to be more clearly differentiated is to study specimens where the top surface layer remains unchanged, while the materials below are varied. The existing literature on layered materials, such as graphene, provide examples of this, where friction has been observed to decrease with increasing layer count.[18] The observed layer dependence is often attributed to puckering effects, which are only expected in materials with weak interlayer forces. Possible contributions from electron-phonon coupling have also been considered[19] although at least one report claims they are negligible.[20] More relevant to the work here is a theoretical study on friction of a layered material, where friction results from the viscous dissipation within the evanescent waves and stress fields excited below the surface by a vibrating tip.[21]

In the work discussed here, we present AFM sliding friction measurements on six $[LaMnO_3]_m/[SrMnO_3]_n$ superlattice films, with varying layer thicknesses in the range 2 to 20 atomic layers (or 0.8 to 8 nm). The films were found to be doped with fluorine, which strongly affected the measured friction coefficients, complicating the systematic investigation of the effect of layer thickness. However, in several films with similar fluorine content, we find indications that top layer thicknesses below around 5 nm affect friction, suggesting that dissipation occurs in the sub-surface material to a depth of 5 nm. We also find that the friction coefficient is consistently constant for a given film, although the friction forces may vary considerably as a function of position due to effects of adhesive forces, suggesting that there may be the possibility to systematically control sliding friction coefficients at manganite surfaces through their composition and surface layer thickness.

## 2. Experimental Section

### 2.1 Specimen Preparation and Characterization

Epitaxial superlattice films of $[LaMnO_3]_m/[SrMnO_3]_n$, where m refers to the number of $LaMnO_3$ (LMO) unit cell layers, and n to the number of $SrMnO_3$ (SMO) unit cell layers, were grown using metal organic aerosol deposition technique (MAD).[22] An overview of the m to n ratio of the films studied here can be found in **Table 1.** Each superlattice stack was terminated with a layer of $[LaMnO_3]_m$ to promote equivalent chemical composition on the surface. The layer thicknesses and periodicities were monitored during deposition using in-situ ellipsometry measurements (see **Section SI-1.1**). The total superlattice film thicknesses were kept constant at around 30 nm. All films were deposited on $5 \times 5 \times 0.5\ \mathrm{mm}^3$ (100)-oriented $SrTiO_3$: 1.0 at. % Nb $K \leq 0.5°$ substrates.

The films were characterized after deposition using standard x-ray methods. $\theta - 2\theta$ diffraction (XRD) confirmed heteroepitaxy and no reflexes related to impurities (see **Section SI-1.2**). Small-angle x-ray scattering (XRR) provides the superlattice reflections which are used to confirm layer thickness and revealed intensities consistent with an interface roughness of RMS $\leq 0.6$ nm for all films (see **Section SI-1.3**).

The elemental composition, chemical state, and electronic state of the film surfaces were obtained using x-ray photoelectron spectroscopy (XPS) (see **Section SI-1.4**). The XPS measurements of the film surfaces showed the expected La, Sr, Mn, and O peaks, as well as peaks for C and F (**Figure 1**). The compositional ratios of (La+Sr)/Mn and Sr/La were estimated from peak intensities using standard methods and found to fall in expected ranges, within the experimental accuracy. The full spectra, as well as further details on the evaluation, are presented in the supporting information (see **Section SI-1.4**).

Fluorine is a well-known source of contamination in materials synthesized in deposition systems containing Teflon components.[25] In our case, the MAD deposition chamber used to grow the superlattice films included Teflon parts at the time of deposition, making this the likely origin of the fluorine found in the films. If fluorine were present solely as a surface contaminant, one would expect a characteristic F 1s binding energy at 689.8 eV, accompanied by a C 1s peak at 292.5 eV, indicating the presence of C–F bonds.[25] However, our XPS measurements reveal a significantly lower F 1s binding energy of 684.45 eV, and no detectable C–F component in the C 1s region at 292.5 eV (see Figure 1**a**). These observations rule out surface contamination as the source of fluorine. Instead, the measured F 1s binding energy is consistent with bulk incorporation of fluorine, such as in $SrF_2$ (684.6 eV)[26] or $LaF_3$ (684.5 eV),[27] suggesting that fluorine substitutes for oxygen within the lattice. This interpretation is supported by previous reports of fluorine doping in $SrMnO_3$, where similar binding energies were observed.[28] Quantitative analysis of the F 1s and Mn 2p peak intensities (see Figure 1**b** and **Section SI-1.4**) indicates that fluorine replaces oxygen at a fractional level between 0.1 and 0.4. The films with top $LaMnO_3$ layer thicknesses of m = 2, 4, and 14 show particularly elevated fluorine concentrations compared to the other superlattice films.

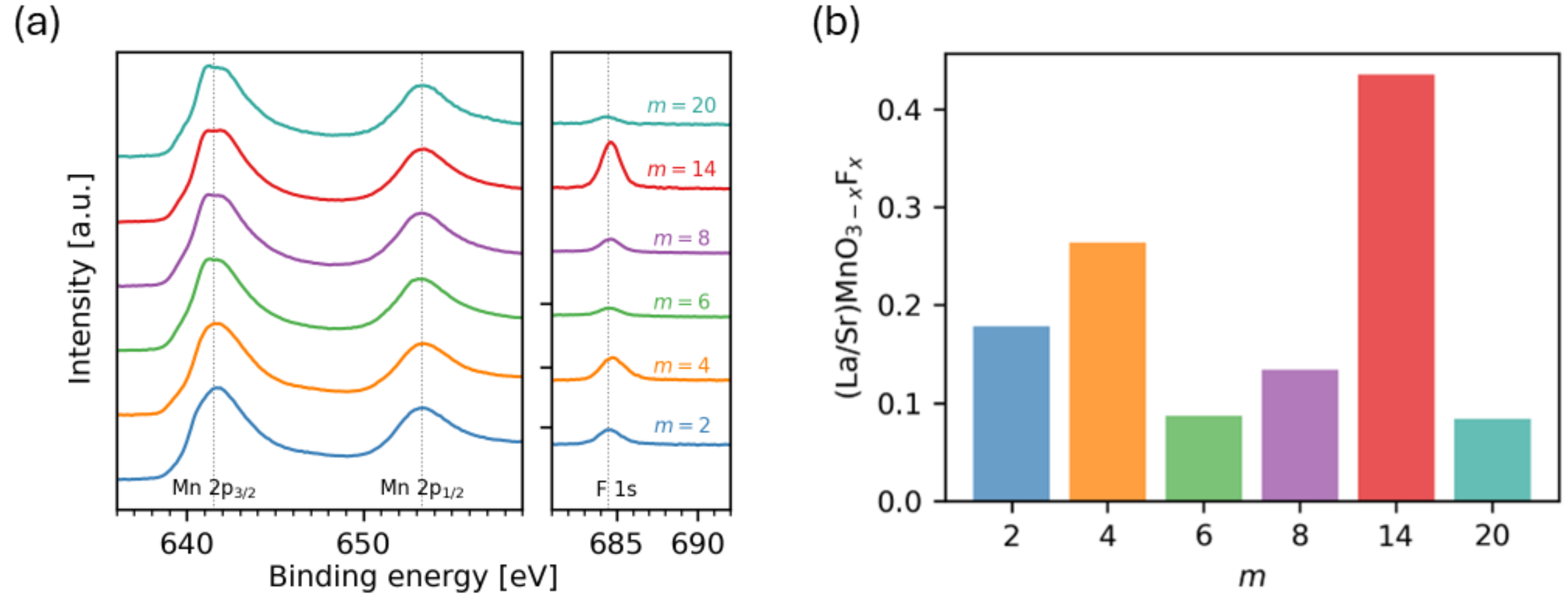

***Figure 1***: *(a) XPS spectra showing the Mn $2p_{1/2}$ and Mn $2p_{3/2}$ peaks at binding energies of 653 eV and 641 eV, and the F 1s peak at a binding energy of 685 eV. (b) The fluorine concentration in the near surface region of the superlattice samples is determined from the ratio of the Mn to F peak areas.*

The thermal resistivities $\kappa^{-1}$ of the films studied here were estimated from prior measurements on superlattice films deposited using identical protocols and in the same deposition chamber.[23] Specifically, $\kappa^{-1}$ values for various $[LaMnO_3]_m/[SrMnO_3]_n$ films, including those with the same m and n ratios as in this study, were obtained via optical transient thermal reflectivity and showed a linear dependence on interface density. This linear relation was used to estimate the $\kappa^{-1}$ values of the films used here, listed in **Table 1** (**see Section SI-1.5**). The comparison is further supported by temperature dependent SQUID measurements (see **Section SI-1.6**), which confirmed that the magnetic properties of the present films align well with those reported earlier.[23,24]

**Table 1.** *$[LaMnO_3]_m/[SrMnO_3]_n$ superlattice film layer thicknesses and thermal resistivities (see **Sections SI-1.3** and **SI-1.5**). $LaMnO_3$ is either cubic ($m/n = 1$) or rhombohedral ($m/n = 2$), depending on the $m/n$ ratio.*

| m [3.9 Å] | n [3.9 Å] | Film thickness [nm] | Thermal resistivity $\kappa^{-1}$ [mK/W] |
|---|---|---|---|
| 2 | 2 | 32 | 0.91(2) |
| 4 | 2 | 34 | 3.27(7) |
| 6 | 6 | 32 | 0.43(2) |
| 8 | 4 | 31 | 1.84(7) |
| 14 | 7 | 32 | 1.23(7) |
| 20 | 10 | 35 | 0.99(7) |

## 2.2 Lateral Force Microscopy

Friction measurements were performed using the lateral force microscopy method in an AFM at room temperature ($T = 293$ K) under UHV ($p \approx 2 \times 10^{-10}$ mbar) conditions. The torsion of the cantilever is measured as it is laterally dragged over the film surface at a constant applied normal force $F_N$ and velocity ($v = 250$ nm $s^{-1}$) and is directly proportional to the lateral forces $F_L$ during sliding. To separate changes in topography from friction effects, so-called friction loops – trace and retrace scanning along the same line on the film surface – are recorded, to yield the friction force $F_F = 1/2 \cdot (F_{L,trace} - F_{L,retrace})$.[29]

Friction forces $F_F$ were obtained by averaging over 50 friction loops that were recorded within a $100 \times 250$ $nm^2$ surface region with a sampling density of 1 $nm^{-1}$ along the fast, and 0.5 $nm^{-1}$ along the slow scan direction for each normal load and film. To assess how robust and reproducible the friction measurements are and to determine the order of magnitude of possible variations, the measurements were repeated several times on randomly selected areas on the film surface. In addition, to minimize wear during the friction experiments, the applied normal forces $F_N$ were kept below 30 nN and any possible changes in the contact were monitored through adhesion measurements.

By measuring the frictional forces on the same surface area as a function of applied load, the friction coefficient μ between the tip and the specimen can be determined using a modified Amontons relation

$$F_F = \mu \cdot F_N + F_{F0}, \tag{1}$$

where the term $F_{F0}$ is the frictional force at $F_N = 0$ N. It is non-zero due to adhesion forces $F_A$ between tip and film, which become comparable to the friction forces at the nanoscale.[29]

Adhesion forces were equated to the pull-off forces obtained by averaging over 50 force-distance curves.[17] Significant changes in adhesion forces before and after probing the frictional properties can indicate changes in tip geometry due to wear, changes in the surface chemistry, or electrostatic forces due to triboelectrification.[30] To confirm that our friction measurements did not alter the surface morphology or affect the friction forces measured, an overview height map and lateral force scan of the measurement area was performed after friction measurements.

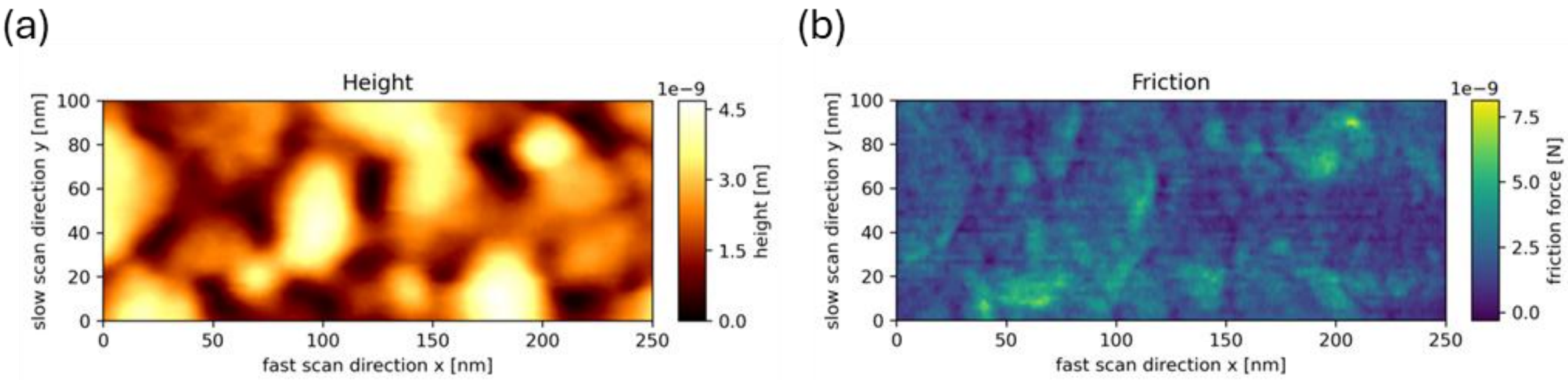

**Figure 2.** *(a) Topography and (b) corresponding friction force $F_F$ maps of a $[LaMnO_3]_2/[SrMnO_3]_2$ surface obtained under a normal load of 28 nN. There is no obvious correlation between the friction force map and the topography.*

### 3. Experimental Results

**Figure 2** shows a height map and the corresponding friction force map for a $[LaMnO_3]_2/[SrMnO_3]_2$ film. The height map has an RMS roughness of less than 0.5 nm and a typical feature size of around 50 nm, while the friction force map shows factor of four variations with feature sizes ranging from 50 nm down to around 5 nm. Spearman correlation coefficients of $r_S = 0.029$ between the friction and height maps and of $r_S = 0.125$ between the friction and height gradients along the x-direction, rule out correlations between friction and topography and support the validity of the lateral force method to measure friction.[29] Similar surface topographies and friction maps were obtained on all $[LaMnO_3]_m/[SrMnO_3]_n$ films and show comparably low correlations $r_S < 0.15$ (see **Section SI-3**).

According to Hertz contact theory, the contact area between the nominally 10 nm radius AFM tip and a flat sample surface should have a contact radius between 0.5 and 1.5 nm, for the range of normal forces experienced here. The characteristic topographical feature size in **Figure 2(a)** is around 50 nm, which suggests that the surface is smooth below this length scale. However, roughness within the contact area at length scales of smaller than 0.5 nm, such as due to atomic scale roughness, cannot be ruled out. In fact, atomic scale roughness would lead to multiple asperities within the contact area and can explain the local variations in friction force observed over length scales as small as 10 nm (**Figure 2(b)**).

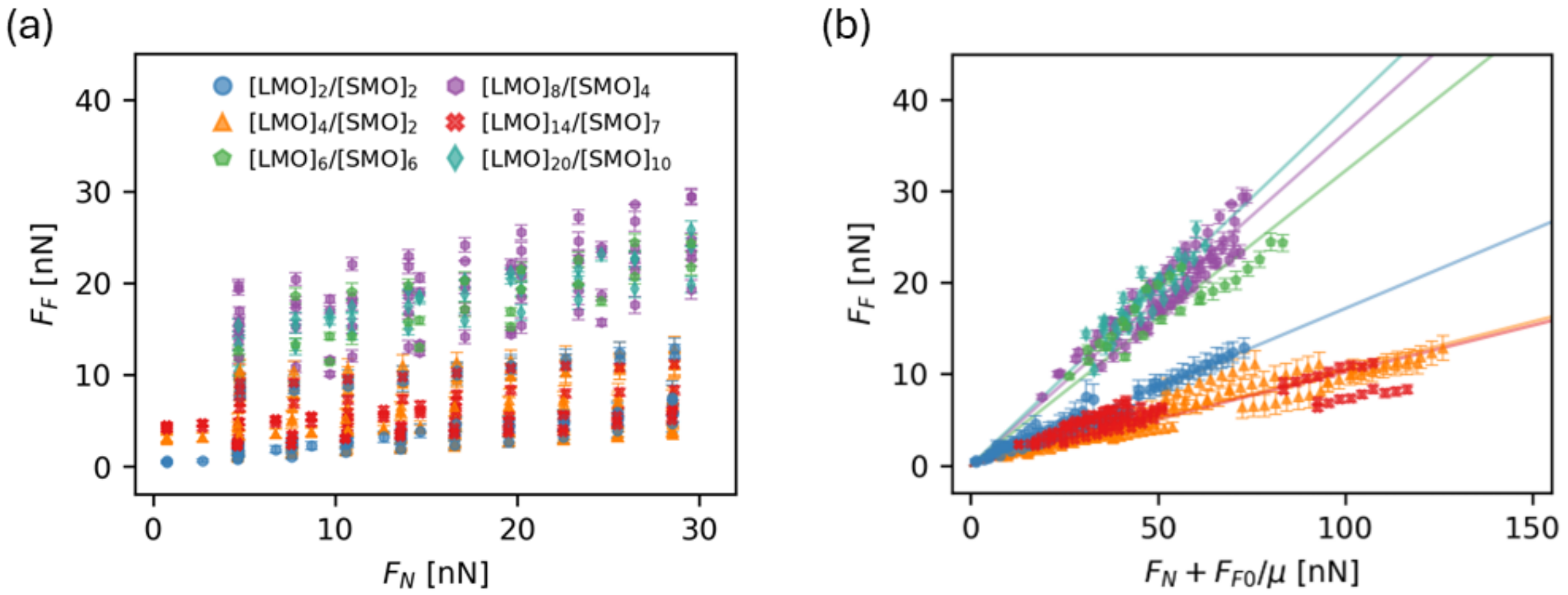

**Figure 3**. *Average friction forces $F_F$ of the $[LaMnO_3]_m/[SrMnO_3]_n$ films plotted as a function of (a) the applied normal load $F_N$ (Amontons measurement), and (b) the total normal load $F_N + F_{F0}/\mu$.*

The friction force $F_F$ for a given film and for a given applied normal load $F_N$ was obtained by averaging over an individual friction map (250 nm × 100 nm region). Sets of friction maps obtained for different applied normal loads between 1 and 30 nN were performed within around a micrometer distance of each other on the sample surface and used to obtain so-called Amontons measurements (friction force versus normal force). These Amontons measurements all show linear behavior, and linear regressions were performed to obtain the non-zero friction force intercept $F_{F0}$ (**Equation 1**) as well as a friction coefficient μ (slope) for each Amontons measurement (**Figure (a)** and **Section SI-4.1**). Several Amontons measurements were obtained for each sample. While the slopes μ did not vary much for a given sample, vertical shifts in data led to wide variations in $F_{F0}$ with no clear dependence on nominal layer thicknesses or other material properties (see **Section SI-4.1**). Shifts as large as 10 nN were found between sets of maps performed at randomly selected regions on a given film surface using a single AFM tip. These are likely due to local changes in surface chemistry, carbon-based surface residues, and adsorbed water. And shifts as large as 50 nN occurred between Amontons measurements performed on a given sample using different AFM cantilevers. These are attributed to changes in tip shape and contact area.

The widely scattered Amontons measurements (**Figure 3(a)**) can be collapsed onto proportional relationships between friction force and the sum of the applied normal load $F_N$ plus $F_{F0}/\mu$ (**Figure (b)** and **Section SI-4.3**). The value $F_{F0}/\mu$ represents the tensile normal force needed to overcome adhesion and separate the surfaces, assuming a linear friction–load dependence. This value is expected to correlate with the pull-off forces from force–distance curves, which also quantify adhesion. Indeed, our measurements show that $F_{F0}/\mu$ and the pull-off forces are comparable (see **Section SI-4.2**), supporting the interpretation that both reflect adhesive contributions to the normal force. The additive effect of applied and adhesive forces is consistent with elastic contact models that include adhesion.[14,31]

Average friction coefficients $\bar{\mu}$ for each film (**Figure 4**) were obtained from the best fit slopes to the data in **Figure (b)**. The friction coefficients $\mu$ and $\bar{\mu}$, which range between 0.1 and 0.4, are comparable to values reported for AFM friction measurements on other manganite thin films.[6,10] They are plotted against the thermal resistivity $\kappa^{-1}$, fluorine content *x*, and top layer thickness d of each film in **Figure 4**. No systematic correlation between friction and thermal resistivity can be seen, indicating that friction is not governed by the GHz–THz phonons responsible for thermal conduction at room temperature.[23] This also suggests that heat dissipation through thermal conduction does not play a dominant role in controlling friction.

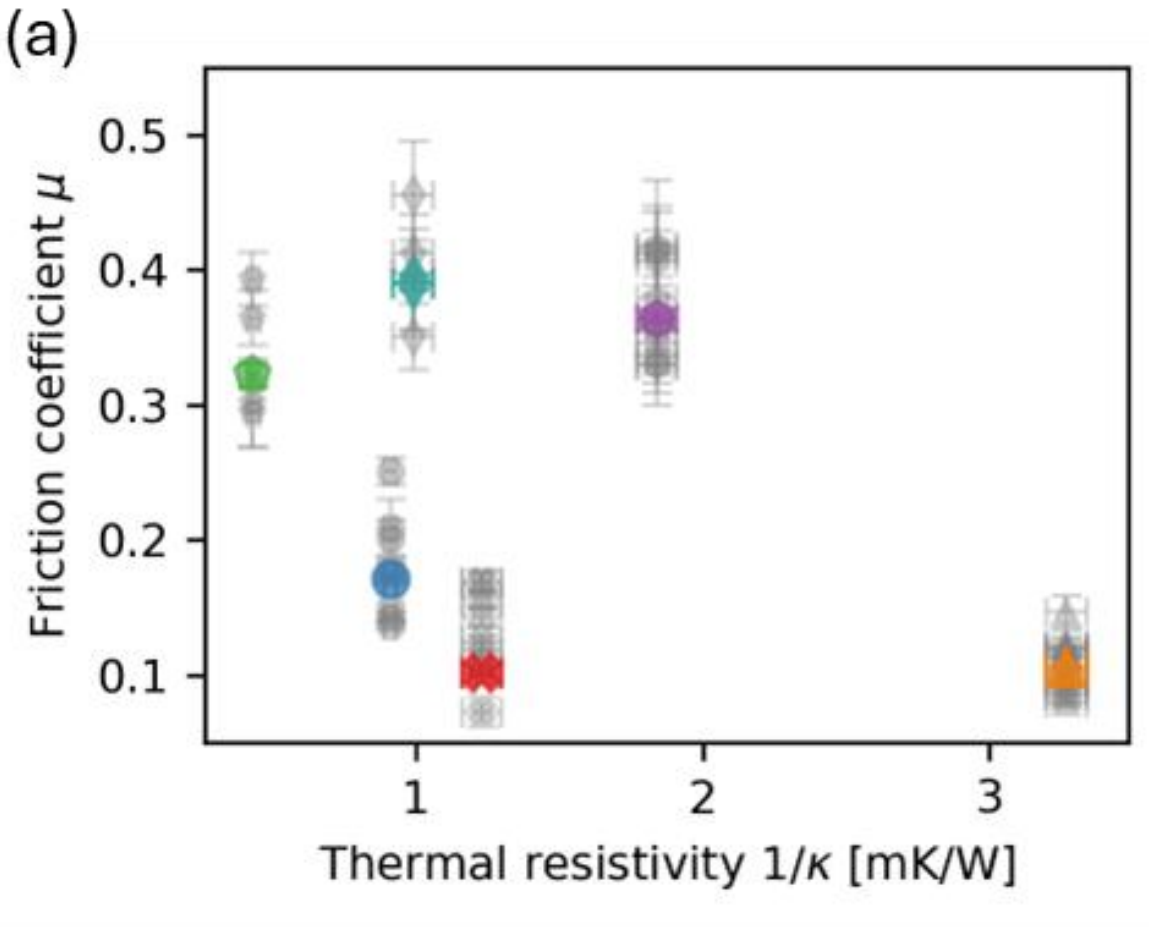

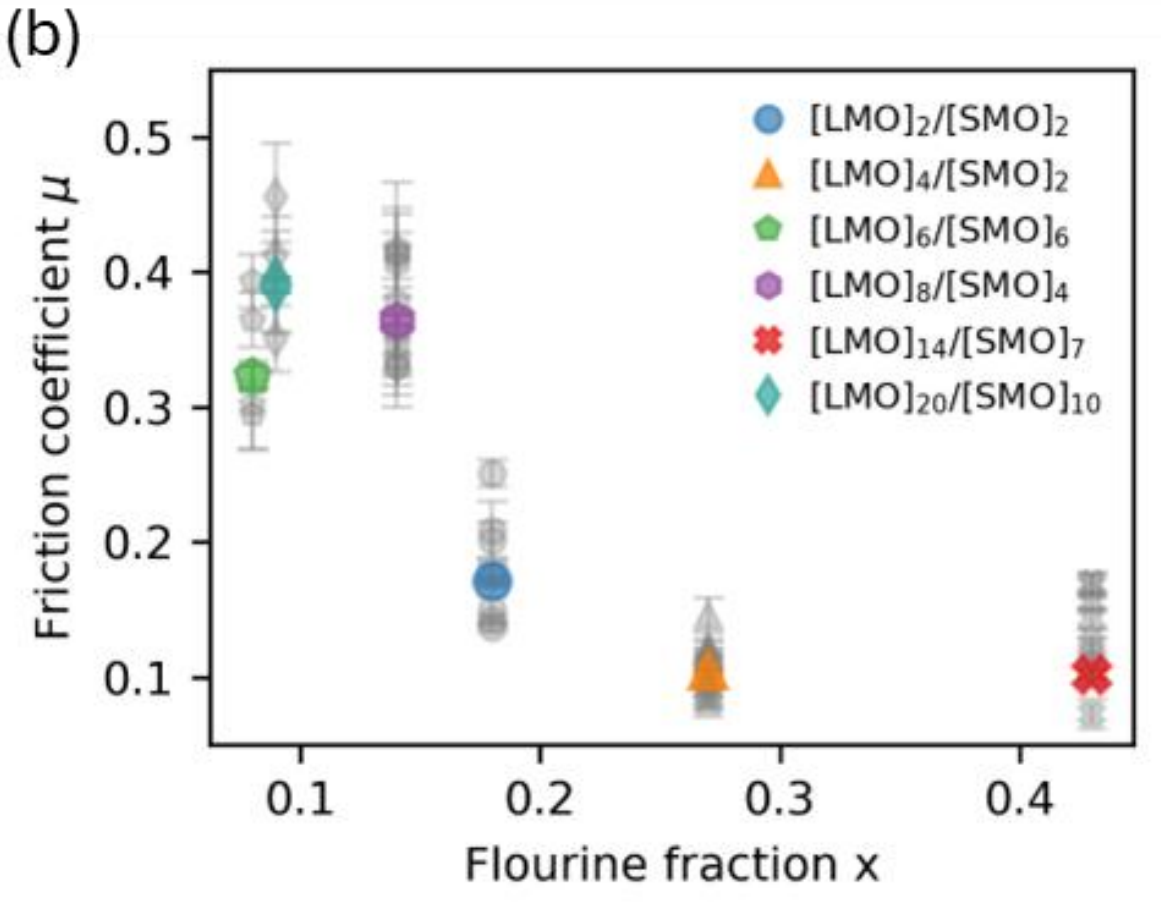

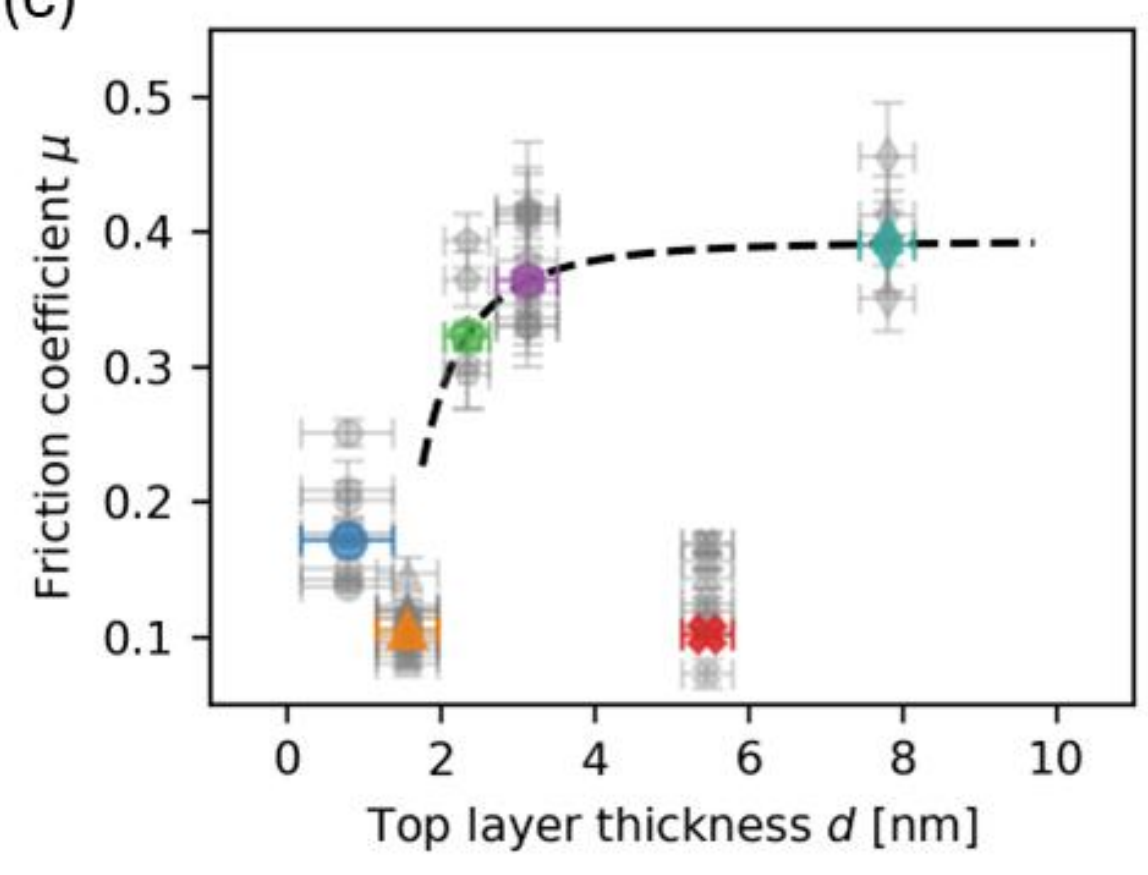

**Figure 4**. *Friction coefficients of the superlattice films plotted against (a) thermal resistivity* $\kappa^{-1}$*, (b) fluorine content x, and (c)* $LaMnO_3$ *top layer thickness d. The friction coefficients* $\mu$ *obtained from each individual Amontons measurement are shown as gray data points (from* ***Figure 3(a)*** *and* ***Section SI-4.1****); the colored data points are the average values* $\bar{\mu}$ *for each film (from* ***Figure 3(b)*** *and* ***Section SI-4.3).*** *There is no apparent trend of friction coefficient with thermal resistivity (a), while the coefficients clearly decrease with increasing fluorine content (b). Among the specimens with low fluorine content, there is some indication that the friction coefficient decreases for thin* $LaMnO_3$ *top layer thicknesses (dashed line) (c).*

In contrast, the friction coefficient is clearly reduced by the presence of fluorine (**Figure 4(b)**). The films with top layer thickness of m=2, 4 and 14 exhibit high F concentrations and reduced friction coefficients compared to the other three films.

Within the set of three specimens containing the same fluorine content ($[LMO]_6/[SMO]_6$, $[LMO]_8/[SMO]_4$, and $[LMO]_{20}/[SMO]_{10}$), the thinnest top layer specimen ($[LMO]_6/[SMO]_6$) has a smaller average friction coefficient than the other two specimens (**Figure 4(c)**). Since the top layers of these three films have roughly the same compositions and similar surface roughness (**Sections SI-1.3** and **SI-3**), the dissipative processes in the sliding contact interface are expected to be the same. Thus, any difference can be attributed to dissipation in the sub-surface material.

In contrast to the friction coefficient, the adhesive forces – either estimated from the pull-off forces or the values $F_{F0}/\mu$ – vary strongly from position to position and show no systematic dependence on thermal conductivity, fluorine content, or layer thickness.

## 4. Discussion

Friction forces were measured using lateral force microscopy on six manganite superlattice films with individual layer thicknesses ranging from 0.8 to 8 nm. The surface topography and frictional behavior remained unchanged during measurements, confirming that the tip–sample contact is elastic within experimental resolution and that wear does not contribute significantly to the measured forces. Adhesion forces were found to vary with position—evident from the vertical shifts in the Amontons plots shown in **Figure 2(a)**—but were unaffected by fluorine doping or superlattice structure. In contrast, the friction coefficient for each specimen remained constant across its surface but varied systematically with both fluorine concentration and top-layer thickness. These trends and their implications are discussed to provide broader insights into mechanisms governing wear-free, sliding-contact frictional losses.

Spatial variations in adhesion are attributed to local differences in surface chemistry and tip–sample contact geometry, both of which are influenced by surface roughness. Variations in roughness and curvature may cause local fluctuations in water adsorption and solvent evaporation under ambient conditions. These variations, in turn, influence surface reactions such as carbon contamination or redox processes,[32] which directly affect adhesive forces. According to the Derjaguin–Muller–Toporov (DMT) model for a single-asperity, elastic–adhesive contact, the adhesive force is given by $2\pi R\Delta\gamma$, where R is the (assumed spherical) tip radius and $\Delta\gamma$ is the change in surface energy per unit area upon contact.[14] The measured pull-off forces (3–75 nN, see **SI-4.2**) correspond to realistic $\Delta\gamma$ values of 0.05-1.20 J/m² for a nominal tip radius of R = 10 nm.[31,33]

Measured friction forces scale linearly with the total normal force, defined as the sum of the applied load $F_N$ and the adhesive force. This supports the interpretation that adhesion contributes additively to the normal load.[31,34] Since friction is generally proportional to the true contact area between surfaces,[35] the observed linear dependence indicates that this area increases linearly with total normal force. In AFM contacts, the true contact area is determined by both surface roughness and the applied normal force. The linear scaling suggests a multi-asperity contact regime between the nominally spherical AFM tip and the nominally flat manganite film. In contrast, single-asperity elastic contact—modeled by Hertzian theory—predicts a sublinear dependence of contact area on normal force, even when adhesion is included.[29] Several multi-asperity contact models, however, predict linear behavior,[29] and both regimes have been observed in AFM studies of hard materials. The specific regime depends on experimental factors such as tip radius, roughness, and applied load. Given the shallow indentation depths (< a few Å) and small contact diameters (~3 nm) in our experiments, atomic-scale surface roughness on either the tip or film is sufficient to induce a multi-asperity contact and account for the observed linear friction–force relationship.

We observe a clear decrease in the friction coefficient with increasing fluorine doping. This is not surprising since fluorine doping is known to alter structural, electrical, and magnetic properties of manganites. In compounds such as $SrMnO_{3-x}F_x$, fluorine doping has been shown to induce ferromagnetism, trigger charge-ordered magnetic transitions through electron doping and local Mn-octahedral distortions,[28] and reduce electrical conductivity.[36] Similarly, fluorine incorporation leads to short-range magnetic ordering in $LaMnO_{2.8}F_{0.2}$ [37] and induces ferroelectricity in $LaMnO_2F$.[38] Despite these substantial changes in bulk properties, fluorine doping does not affect adhesion forces, in agreement with XPS binding energy data indicating that fluorine is not a surface contaminant.

Among superlattice films with similar fluorine content, the sample with the thinnest $LaMnO_3$ top layer (2.3 nm) exhibits a 15% lower friction coefficient than films with thicker top layers. This suggests that material properties within ~5 nm of the surface influence sliding friction. The reduction may stem from the underlying $SrMnO_3$ or the LMO/SMO interface when positioned close to the surface. It is unclear whether the effect arises from intrinsic differences between LMO and SMO or from the interface itself. Although LMO has a ~7% lower elastic modulus than SMO,[39,40] Hertzian contact mechanics predict less than a 1% change in contact area across the thickness range. Thus, the reduced friction in the $[LMO]_6/[SMO]_6$ film is more likely due to decreased energy dissipation rather than contact area changes.

We propose that the dominant contribution to the measured friction arises from the generation and dissipation of mechanical excitations—such as stress fields, vibrations, and non-thermal phonons.[6] These excitations are primarily induced by Pauli repulsion forces and extend well beyond the immediate contact interface. For instance, recent theoretical studies modeling an oscillating AFM tip have shown that evanescent waves and slowly moving stress fields dominate under AFM conditions and extend several nanometers below the surface. Similarly, simulations and analyses have demonstrated that the stress fields generated by an AFM tip in contact penetrate several nanometers into the sample.[35]

The idea that mechanical dissipation via internal friction contributes to sliding friction has been widely explored in the context of elastomers. Internal friction is commonly quantified by the inverse quality factor, $Q^{-1}$, and is often measured using mechanical spectroscopy. In elastomers—especially rubber—it is well-established that sliding friction scales with internal friction or elastic hysteresis losses.[15,16] Various theoretical models support this idea, proposing a linear relation of the form $\mu_{IF} = C \cdot Q^{-1}$, where $\mu_{IF}$ is the friction coefficient due to internal friction, and C is a dimensionless proportionality constant.[17,41,42] This model explains energy dissipation from repeated loading and unloading of the material beneath the sliding asperity, and is consistent with observations in materials ranging from elastomers to metals.[15]

The proportionality constant C captures the mechanical and geometrical specifics of the contact, such as roughness, tip radius, and adhesion. For rubber sliding against rough surfaces, C is typically of order unity.[17] In the case of a single AFM tip with spherical radius R and circular contact radius a, the proportionality constant has been related to the local slope of the surface under the tip, estimated as $a/R \approx 0.1$ based on Hertzian contact mechanics. However, this number should be considered a very rough estimate, as smooth, continuous sliding at the nanoscale is likely an oversimplification.

In reality, mechanical perturbations will affect the tip-sample interaction. Vibrations of the tip or sample, as well as dynamic processes at the contact interface, can result in partial or full unloading events. For example, if atomic-scale stick-slip occurs, the loading/unloading rate can increase substantially—up to $v_s/\lambda$ where $v_s$ is the tip sliding speed and $\lambda$ is the atomic-scale slip distance. This is expected to lead to an increase in dissipated energy by a factor of $2a/\lambda \approx 30$ compared to smooth sliding. However, mechanical vibrations with frequencies up to the cantilever resonance frequency (~100 kHz) are expected to be present in the sliding interface. Such dissipation via loading and unloading events turns up in microscopic models as evanescent waves, which decay exponentially into the material away from the surface.[21] Therefore, the tip-sample contact likely experiences mechanical excitations over a broad frequency range up to ~100 kHz, allowing values up to $10^5$ for the proportionality constant C.

Although the value of C is challenging to determine, it should remain constant for a given contact configuration. From material property charts, $Q^{-1}$ values range from ~$10^{-5}$ for materials like carbides and silicate glasses to ~1 for elastomers.[43] Sliding friction coefficients for macro-scale dry contacts typically fall between $10^{-3}$ and 1, implying practical values of C between 1 and 100, which are reasonable given the above discussion.

For the specific case of the manganites, mechanical spectroscopy studies have reported $Q^{-1}$ values between $10^{-4}$ and $10^{-2}$.[44–46] These spectra show a baseline loss from non-activated processes such as thermoelastic strain-temperature coupling, superimposed with broad peaks attributed to magnetic, structural, and charge-ordering transitions. These features arise from processes involving characteristic excitation and relaxation times. Notably, manganites exhibit higher internal friction than many other technologically relevant ceramics,[43] likely due to strong coupling between strain, spin, and charge degrees of freedom. This enhanced internal friction may explain the observed changes in sliding friction across phase transitions in several manganite systems.[6,10] In any case, the range of reported $Q^{-1}$ values is consistent with the friction coefficients measured in our experiments (0.1–0.4), supporting the idea that dissipation of mechanical excitations via internal friction is a viable explanation for the energy loss during AFM tip sliding.

Collectively, our findings provide a coherent picture of tip–sample interactions during AFM friction measurements on manganite superlattice films. The data reveal an elastic, adhesive, multi-asperity contact in which both contact area and friction force scale linearly with the total normal force. This behavior is characteristic of non-wear friction regimes and suggests that the underlying mechanisms may be broadly relevant beyond manganites.

Importantly, the adhesion and friction coefficients exhibit independent trends, indicating distinct origins: adhesion is primarily surface-controlled, while friction is governed by subsurface dissipation extending several nanometers below the surface. Our results further show that friction can be tuned by adjusting the thickness of surface layers. This concept extends beyond sliding contacts; for example, theoretical studies of non-contact AFM tips oscillating above coated substrates have found similar dependencies of energy dissipation on top-layer thickness.[21] Under typical AFM conditions, the tip induces evanescent deformation waves that penetrate the near-surface region to depths comparable to the vibration amplitude. The dominant energy loss arises from internal friction within this excited volume, supplemented by dissipation carried away through volume and surface waves.

## 4. Conclusion

The AFM-based friction measurements of $[LaMnO_3]_m/[SrMnO_3]_n$ superlattice films demonstrate that frictional behavior can be understood within a framework of multi-asperity elastic contact, where energy dissipation is quantified by a robust and reproducible friction coefficient. This coefficient is largely insensitive to surface conditions but strongly dependent on subsurface material properties. We propose that the dominant dissipation mechanisms are closely related to those responsible for internal friction, offering a broadly applicable framework for understanding friction and selecting materials with tunable frictional behavior.

Our interpretation—that subsurface energy dissipation via internal friction plays a dominant role in manganites—suggests new strategies for friction control in a wide range of materials and at multiple length scales. Since macroscale frictional contacts consist of numerous nano- to microscale asperities, similar subsurface mechanisms are likely to operate. In such systems, materials with higher internal friction may offer greater potential for friction tuning through deliberate modifications of subsurface structure.

## Supporting Information

The Supporting Information is available free of charge at
https://pubs.acs.org/doi/10.1021/acsami.xxxxx.

## Acknowledgements

This work was funded by the German Research Foundation (DFG) 217133147/SFB 1073, Project A01 and Project A02. The authors thank D.R. Baer, M.H. Engelhard from Pacific Northwest National Laboratory for discussion and help in interpreting the fluorine contamination observed in XPS measurements. We thank T. Brede for x-ray characterization of all films and SEM measurements of cantilevers. Additionally, we thank A. Wodtke and F. Güthoff from the Max

Planck Institute for Multidisciplinary Sciences Göttingen for providing access and to maintain the Omicron VT-AFM system. We also thank J.M. Rybak for assistance during the preparation and submission of the manuscript.

## Supporting Information

# Nanoscale friction of manganite superlattice films controlled by layer thickness and fluorine content

Niklas A. Weber[a], Miru Lee[b], Florian Schönewald[a], Leonard Schüler[c], Vasily Moshnyaga[c], Matthias Krüger[b], and Cynthia A. Volkert[a,d,e,*]

[a] Institute of Materials Physics, University of Göttingen, Friedrich-Hund-Platz 1, Göttingen 37077, Germany

[b] Institute for Theoretical Physics, University of Göttingen, Friedrich-Hund-Platz 1, 37077 Göttingen, Germany

[c] 1st Physics Institute, University of Göttingen, Friedrich-Hund-Platz 1, 37077 Göttingen, Germany

[d] The International Center for Advanced Studies of Energy Conversion (ICASEC), University of Göttingen, Göttingen 37077, Germany

[e] International Institute for Carbon Neutral Energy Research (WPI-I2CNER), Kyushu University, 744 Motooka, Nishi-ku, Fukuoka 819-0935, Japan

* Email: cynthia.volkert@phys.uni-goettingen.de

## SI-1 Material Characterization

### SI-1.1 Ellipsometry Measurements

The thicknesses of the superlattice layers were monitored and controlled during metal-organic aerosol deposition (MAD) using in-situ ellipsometry. The ellipsometric phase shift angle Δ was recorded during the growth of the superlattice films, up to a total thickness of approximately 30 nm (see Figure S1). Based on previously established deposition rates[1]—$v_{LMO}$ = 0.37 u.c./s and $v_{SMO}$ = 0.25 u.c./s—superlattices with the desired $LaMnO_3$ to $SrMnO_3$ $(m:n)$ ratios were fabricated. The resulting layer thicknesses were further verified by X-ray reflectivity (XRR) measurements (see Section SI-1.4).

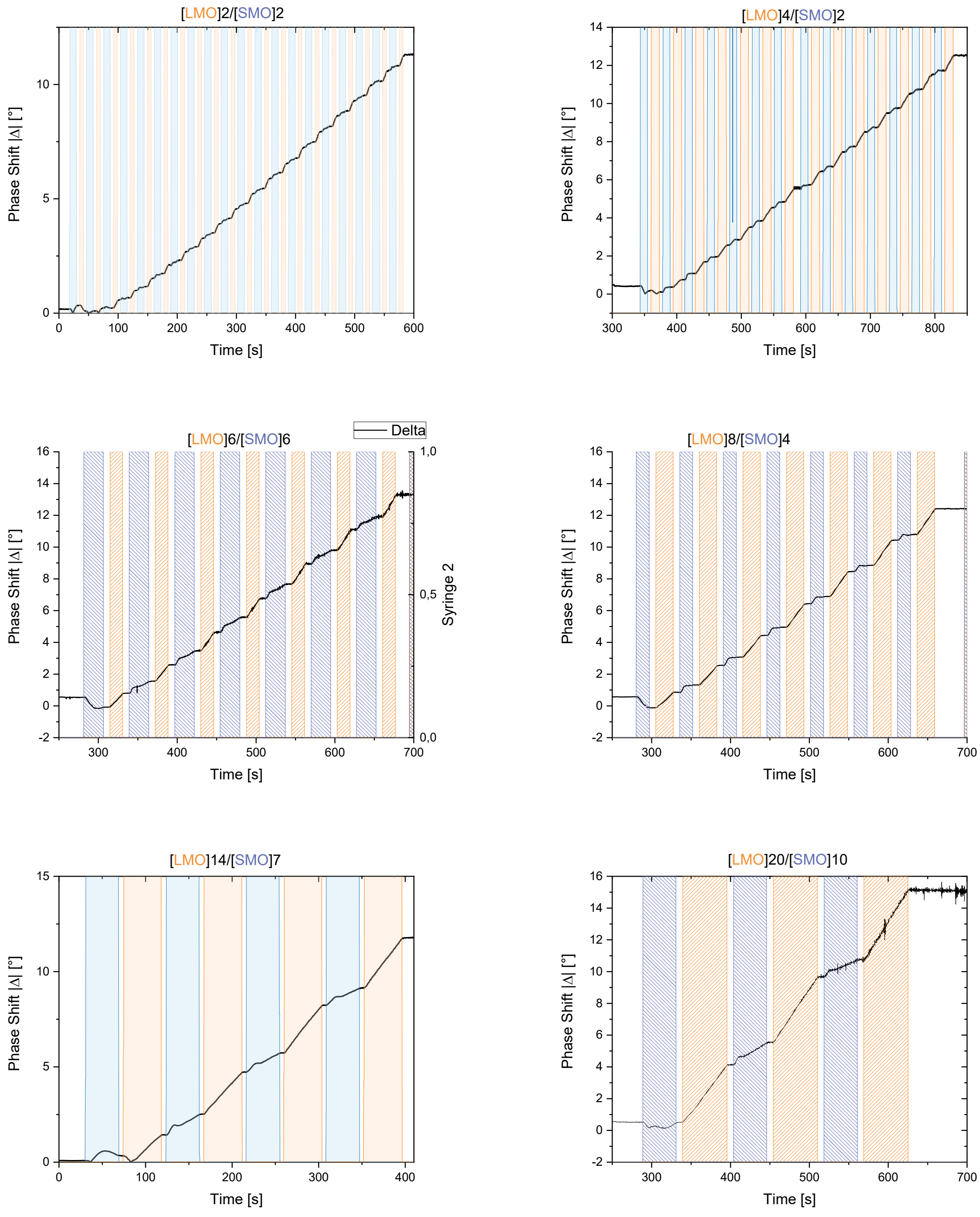

*Figure S1: In-situ ellipsometric phase shift angle (Δ) recorded during the deposition of representative superlattice films. Red-shaded regions correspond to the deposition of $LaMnO_3$ layers, and blue-shaded regions correspond to $SrMnO_3$ layers.*

## SI-1.2 XRD Measurements

X-ray diffraction (XRD) measurements in the Θ–2Θ geometry were carried out using a Bruker D8 Discover diffractometer equipped with a Cu $K_\alpha$ radiation source. The data confirm the presence of superlattice periodicity, as evidenced by the appearance of superstructure reflections. Additionally, weak diffraction peaks attributable to tungsten radiation are observed. Figure S2 presents representative diffractograms from each film.

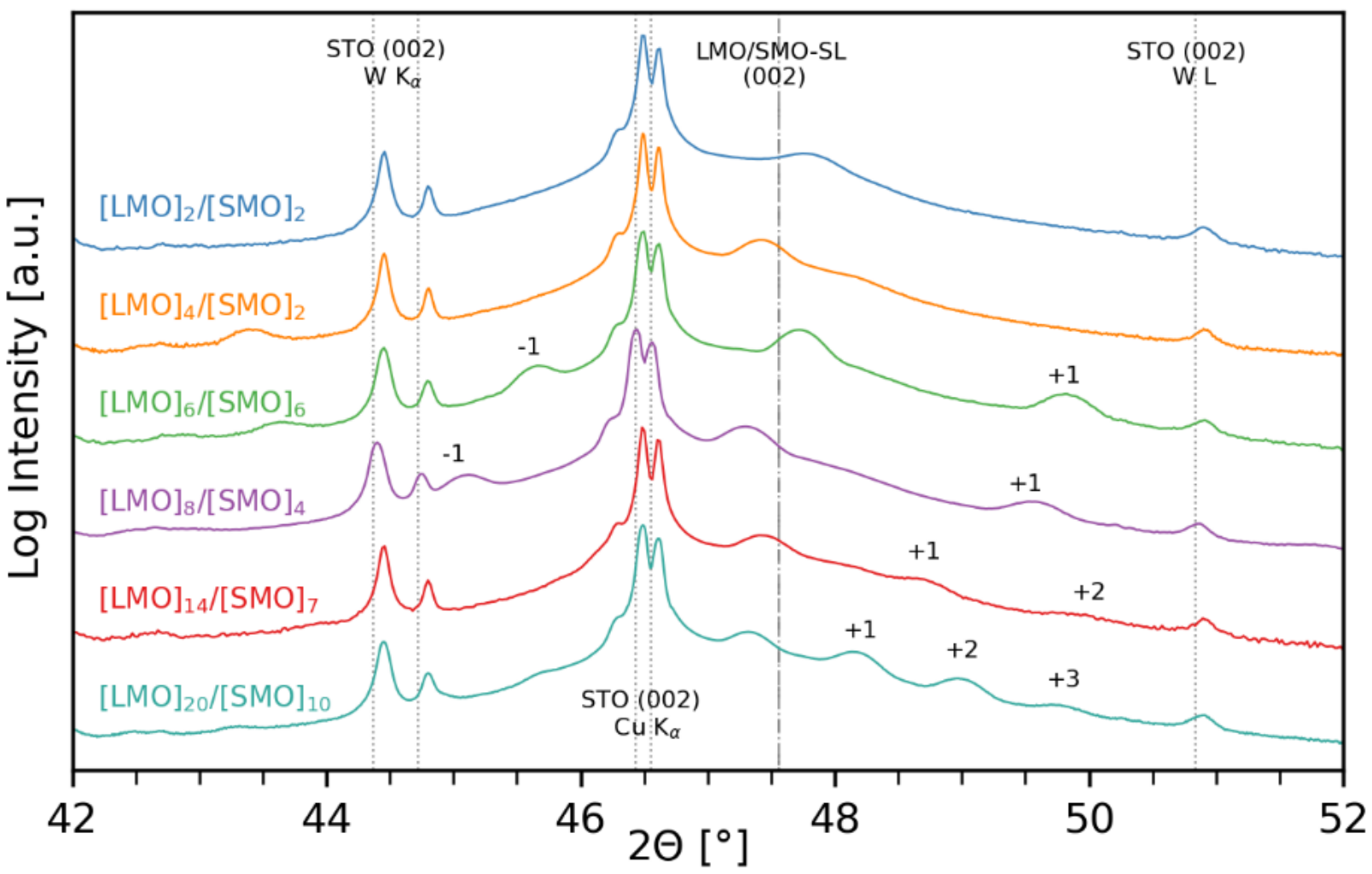

*Figure S2: Θ–2Θ X-ray diffractograms showing characteristic peaks from the $SrTiO_3$ substrate (STO), diffraction peaks from the $LaMnO_3$–$SrMnO_3$ superlattice films (LMO–SMO SL), and superstructure reflections, which are labeled according to their reflection order.*

## SI-1.3 XRR Measurements

XRR measurements were performed using a Bruker D8 Advance diffractometer equipped with a Cu $K_\alpha$ radiation source. Representative reflectivity curves and corresponding fits for all samples are shown in Figure S3. The fits (black lines) were generated using the GenX software package,[2] modeling the samples as a periodic stack of alternating $LaMnO_3$ and $SrMnO_3$ layers with respective thicknesses ($d_{LMO}$, $d_{SMO}$) and interfacial roughness values ($\sigma_{LMO}$, $\sigma_{SMO}$). The best-fit parameters are summarized in Table S1.

For the 2/2 and 4/2 superlattice films, only the total bilayer thickness $d_{LMO} + d_{SMO}$ is listed for the 2/2 and 4/2 films, because the ratio $d_{LMO}/d_{SMO}$ is reported, as the individual layer thickness ratio is primarily determined by the relative intensities of the superstructure reflections. In these films, only the first-order superstructure peak was resolvable due to limitations from interfacial roughness and instrument resolution.[3]

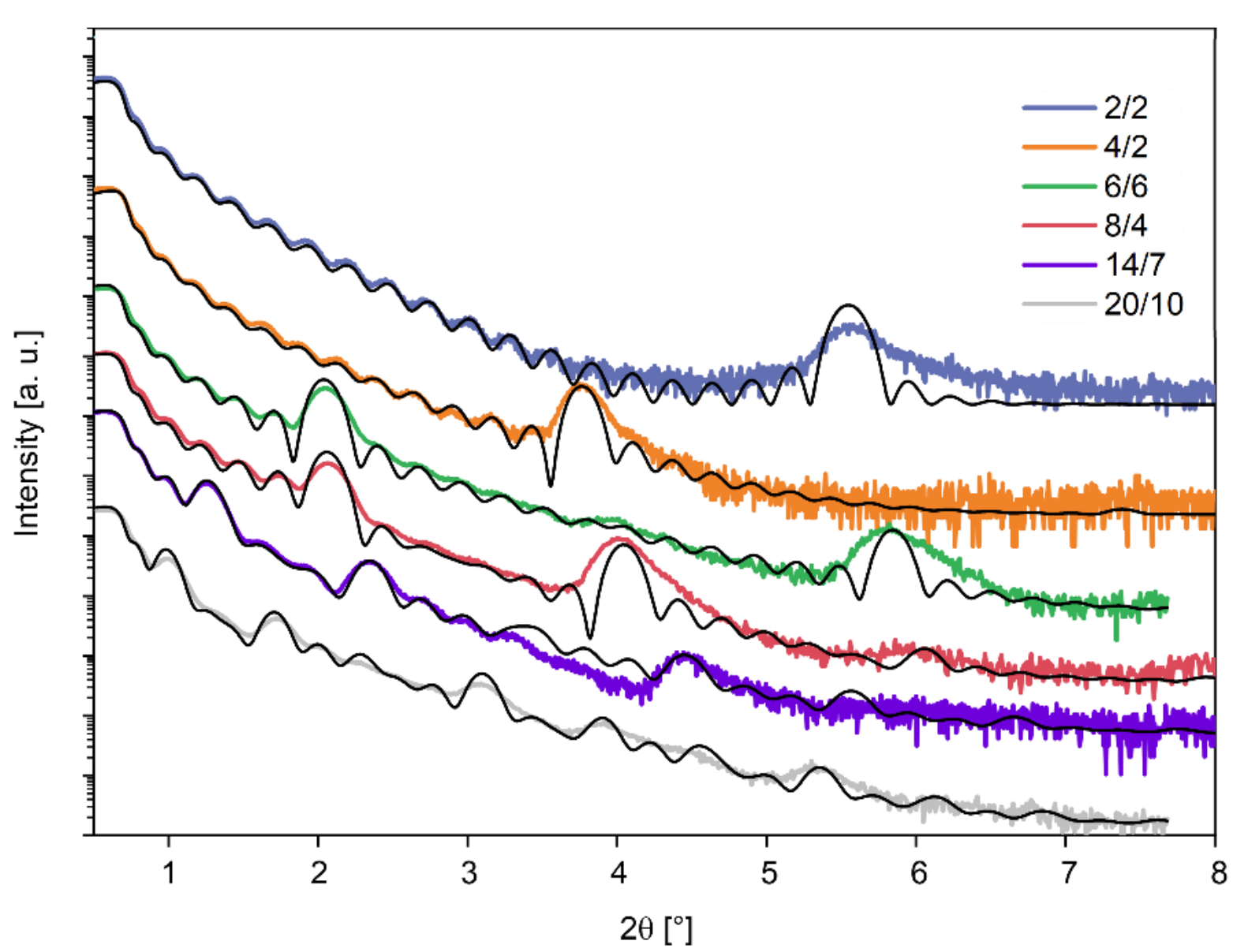

*Figure S3: X-ray reflectivity measurements (colored curves) and model fits (black lines) for the superlattice films. The fits are based on layer thicknesses and roughness parameters listed in Table S1.*

*Table S1: Layer thicknesses and layer roughnesses obtained from the fits to the XRR data. Table S1: Structural parameters extracted from XRR fitting: $LaMnO_3$ and $SrMnO_3$ layer thicknesses ($d_{LMO}$, $d_{SMO}$) and interfacial roughnesses ($\sigma_{LMO}$, $\sigma_{SMO}$).*

| ***m/n*** | $d_{Bilayer}$ **[Å]** | $\sigma_{LMO}$ **[Å]** | $\sigma_{SMO}$ **[Å]** | |
|---|---|---|---|---|
| 2/2 | 16(1) | 6 (1) | 3(2) | |
| 4/2 | 24(1) | 4(1) | 7(3) | |
| ***m/n*** | $d_{LMO}$ **[Å]** | $\sigma_{LMO}$ **[Å]** | $d_{SMO}$ **[Å]** | $\sigma_{SMO}$ **[Å]** |
| 6/6 | 23(1) | 3.1(3) | 23(1) | 3.4(7) |
| 8/4 | 31(2) | 3.8(6) | 13(1) | 4.5(2) |
| 14/7 | 54(1) | 3.3(3) | 27(2) | 6.7(9) |
| 20/10 | 75(2) | 3.6(2) | 42(2) | 5.6(7) |

## SI-1.4 XPS Measurements

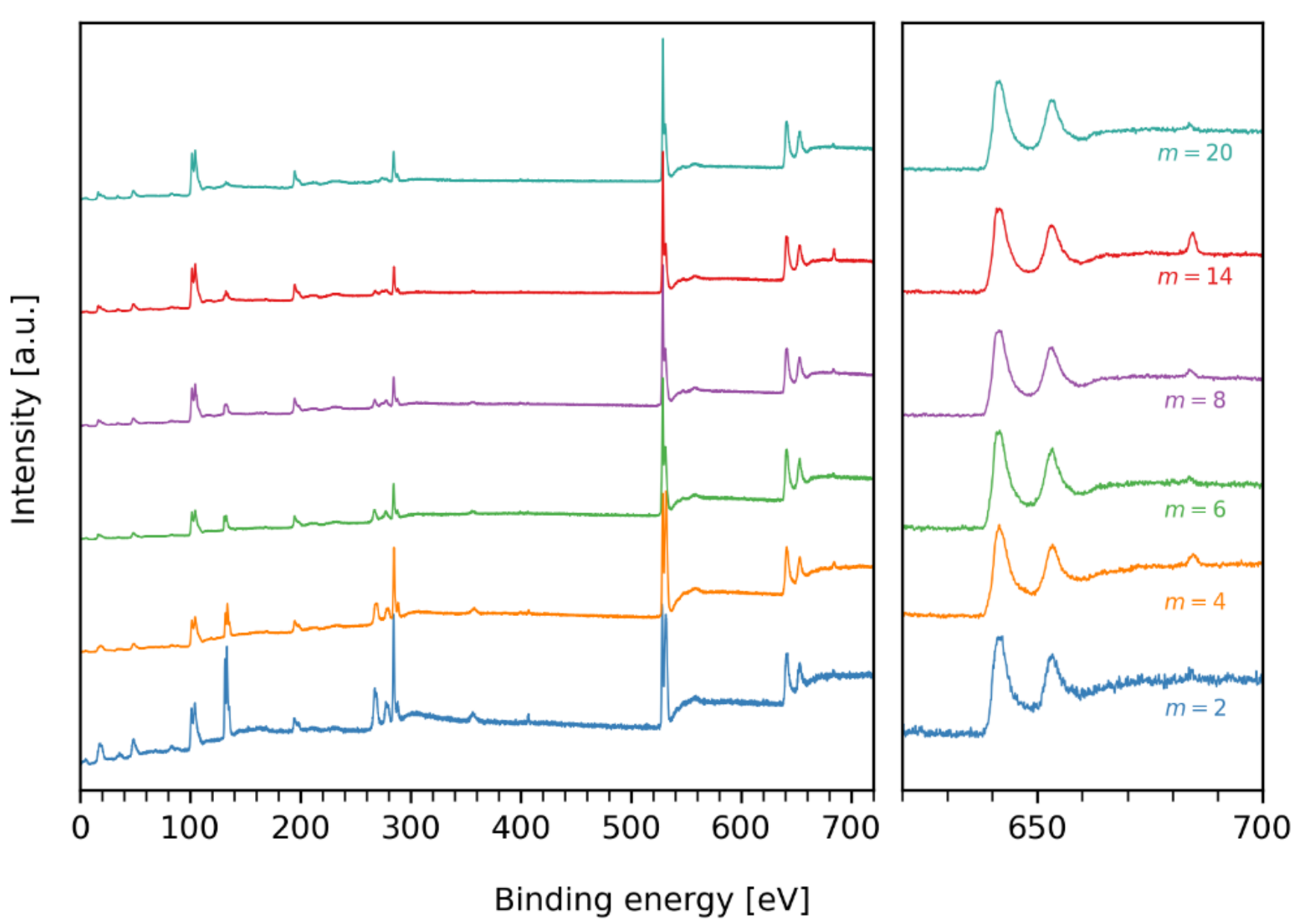

*Figure S4: XPS spectra of representative superlattice films. Left: Overview spectra acquired across the full binding energy range (750–0 eV). Right: High-resolution spectra showing the Mn 2p and F 1s peaks. The binding energy scale has been corrected by setting the main C 1s peak to 284.8 eV.*

X-rayPS measurements were performed on all superlattice films using a Kratos Axis Supra spectrometer equipped with a monochromatic Al Kα X-ray source. Selected portions of the spectra are shown in Figure S4. The overview spectra (left panel) covering the binding energy range from 750 eV to 0 eV were acquired using a pass energy of 40 eV and a step size of 0.1 eV. High-resolution scans of the F 1s, Mn 2p, Mn 3p, and C 1s regions were acquired with a pass energy of 20 eV and the same step size. The Mn 2p and F 1s peaks are presented in the right panel of Figure S4. The binding energy scale was calibrated by referencing the main C 1s peak to 284.8 eV.

Elemental ratios were calculated from the integrated peak intensities using the Escape software, applying instrument-specific relative sensitivity factors (RSFs). Background subtraction was performed using a Shirley-type model, and the data were corrected for the instrument's transmission function.

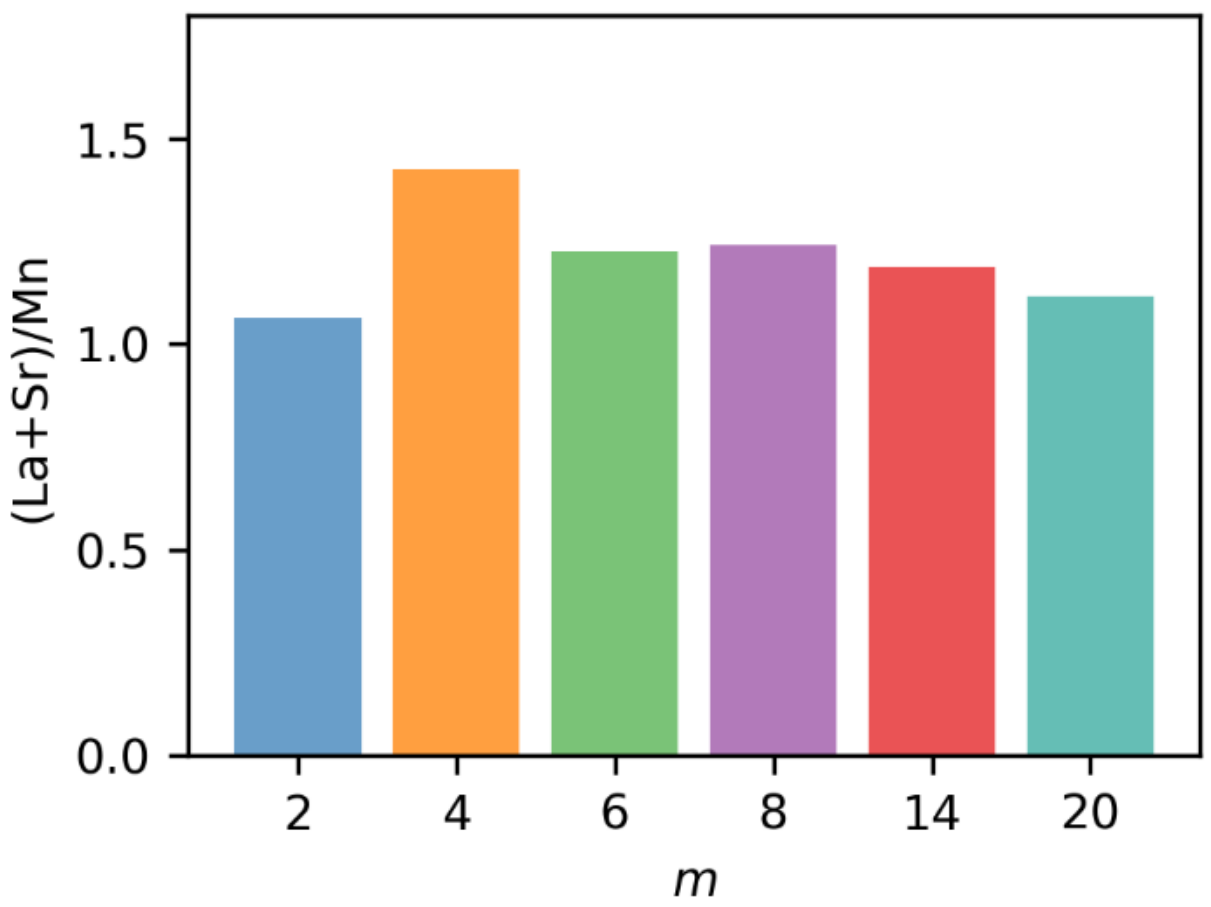

*Figure S5: Elemental ratio of La 4d + Sr 3d to Mn 3p photoemission intensities for all superlattice films.*

The ratio (La + Sr)/Mn is expected to be 1.0 based on the stoichiometry of the superlattices. Mn concentration was estimated using the Mn 3p peak (49 eV), as it lies close in energy to the La and Sr peaks (La 4d at ~102 eV and Sr 3d at ~136 eV), minimizing artifacts due to variations in electron escape depth. The measured ratios range from approximately 1.0 to 1.5, reflecting the expected uncertainties of quantitative XPS analysis.

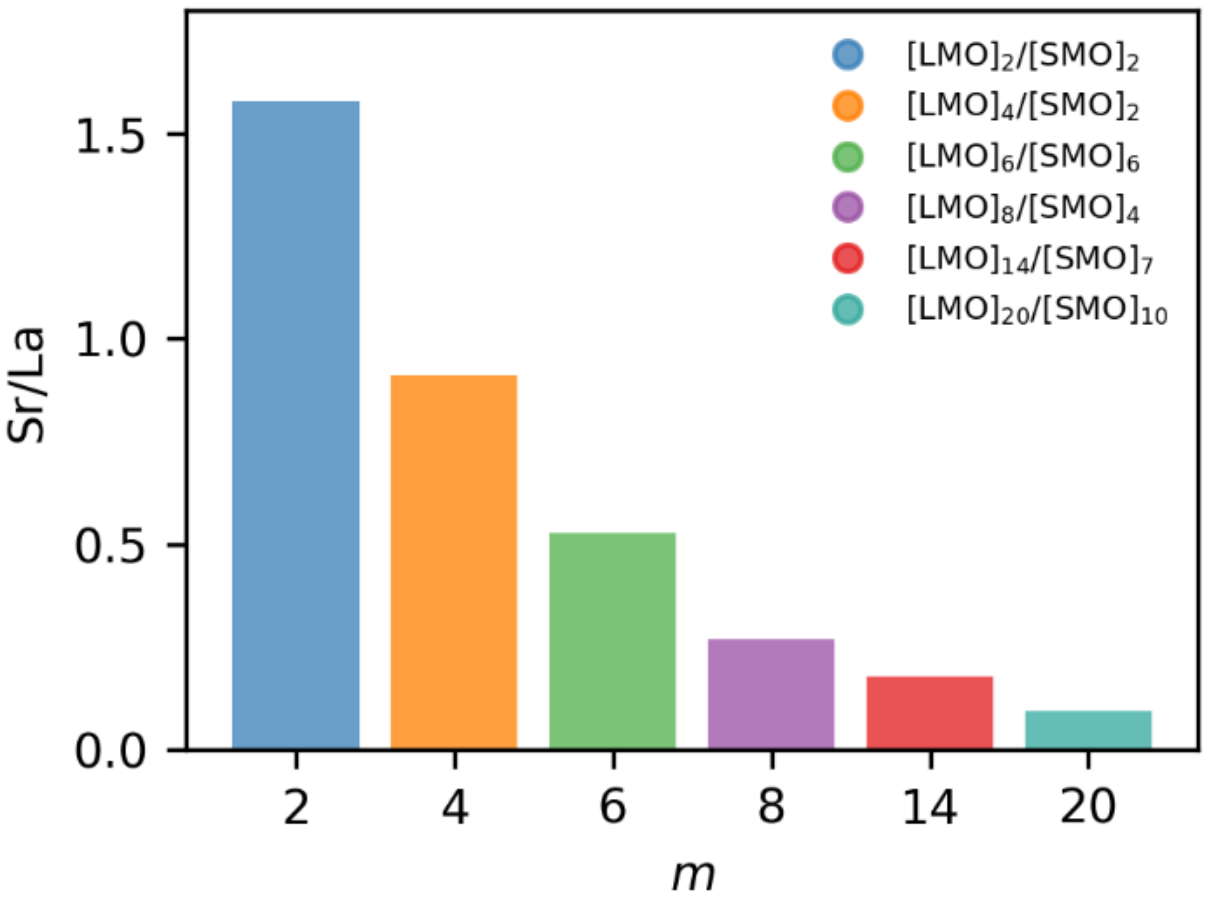

*Figure S6: Measured Sr/La intensity ratio as a function of $LaMnO_3$ top layer thickness.*

The relative Sr signal decreases with increasing $LaMnO_3$ top layer thickness, consistent with the exponential attenuation of photoelectron signal intensity with depth. An inelastic mean free path of $\lambda = 2.3$ nm is assumed for the multilayer system.[4,5] The elevated Sr/La ratio (1.57) observed for the m = 2 sample is likely due in part to significant interface roughness, which approaches the nominal layer thickness (see Section SI-1.3).

### SI-1.5 Thermal Resistivity Values

The thermal properties of superlattice films were characterized using thermal transient reflectometry (TTR). These measurements were conducted on films prepared under similar conditions and with comparable $LaMnO_3$:$SrMnO_3$ ($m/n$) ratios to those used in the present study.[6,7] Measured thermal resistivities for two sets of superlattices—with $m/n$ ratios of 1 (blue triangles) and 2 (orange circles)—are plotted in Figure S7. The $m/n$ ratio can be converted into a unitless interface density using the expression $c/\Lambda$, where $c$ is the unit cell thickness and $\Lambda = m + n$ is the total superlattice period. The estimated thermal resistivities of the films investigated in this work are indicated by green diamonds and listed in the accompanying table.

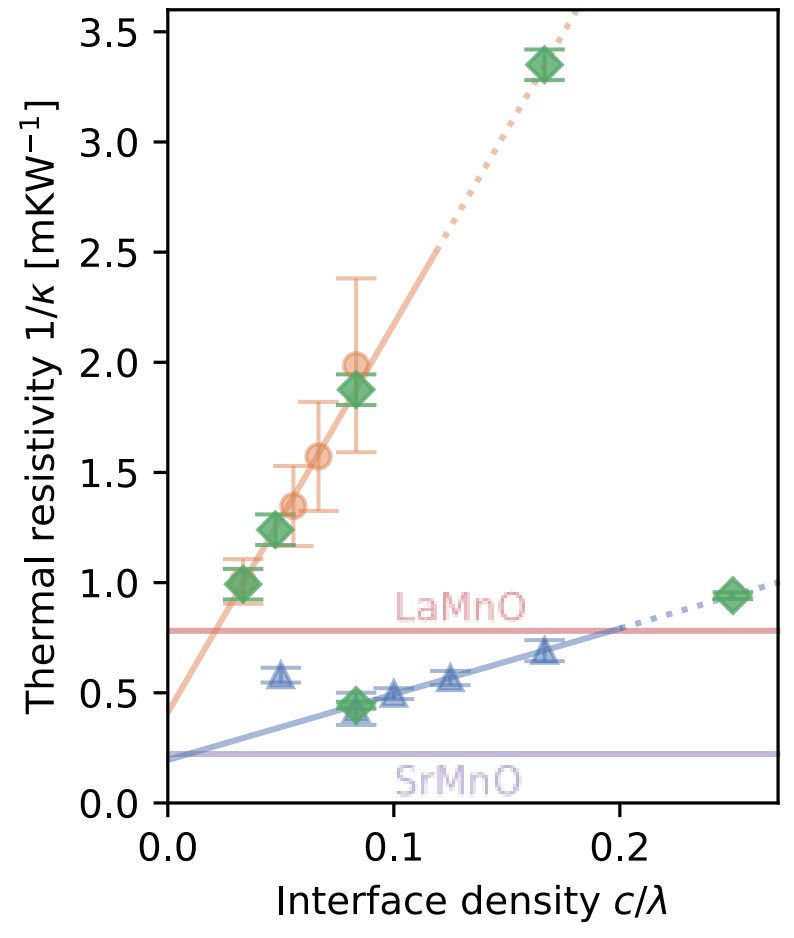

| $c/\Lambda$ | $m$ | $n$ | $1/\kappa$ (mKW$^{-1}$) |
|---|---|---|---|
| 0.25 | 2 | 2 | 0.91(2) |
| 0.166 | 4 | 2 | 3.27(7) |
| 0.083 | 6 | 6 | 0.43(2) |
| 0.083 | 8 | 4 | 1.84(7) |
| 0.048 | 14 | 7 | 1.23(7) |
| 0.033 | 20 | 10 | 0.99(7) |

*Figure S7: Thermal resistivity of superlattice films as a function of interface density (c/Λ), measured by thermal transient reflectometry (TTR). Data for superlattices with $m/n$ = 1 (blue triangles) and $m/n$ = 2 (orange circles) are shown. Green diamonds indicate estimated thermal resistivities of the films used in this study, based on their corresponding interface densities, and are also listed in the accompanying table. Trend lines correspond to linear fits for each $m/n$ group. See Reference 7 for further discussion of the linear model and experimental methodology.*

### SI-1.6 Magnetic Characterization

To verify stoichiometry and probe interfacial charge transfer effects, magnetic measurements were performed using a Quantum Design MPMS-XL SQUID magnetometer. Temperature-dependent magnetization curves for the superlattice and reference films are shown in Figure S8. A 32 nm-thick $LaMnO_3$ reference film exhibits ferromagnetic behavior with a Curie temperature $T_C$ = 152 K, consistent with expectations for stoichiometric $LaMnO_3$ under compressive epitaxial strain from the $SrTiO_3$:Nb (1.0 at.%) substrate.[8]

For the superlattice films, charge transfer at the $[LaMnO_3]_m/[SrMnO_3]_n$ interfaces—extending up to approximately three unit cells—leads to magnetic properties comparable to those of disordered $La_{m/(m+n)}Sr_{n/(m+n)}MnO_3$ alloys. For instance, the $[LaMnO_3]_{2n}/[SrMnO_3]_n$ superlattices display characteristics of optimally doped $La_{2/3}Sr_{1/3}MnO_3$ ($T_C \approx 350$ K), while the $[LaMnO_3]_n/[SrMnO_3]_n$ series exhibits behavior similar to $La_{0.5}Sr_{0.5}MnO_3$, an A-type antiferromagnet.[9] With increasing individual layer thickness, the magnetic properties gradually evolve toward those of bulk $LaMnO_3$ (ferromagnetic) and $SrMnO_3$ (antiferromagnetic), evidenced by a systematic reduction in saturation magnetization and an increase in coercive field, as shown in Figures S9 and S10. These trends suggest smooth, well-defined interfaces with minimal intermixing.

Overall, the observed magnetic properties are consistent with previous reports on metal-organic aerosol deposition (MAD)–grown superlattices and confirm the presence of a high-$T_C$ interfacial magnetic phase, thereby supporting the validity of using thermal resistivity data from D. Meyer et al.[7] in this study.

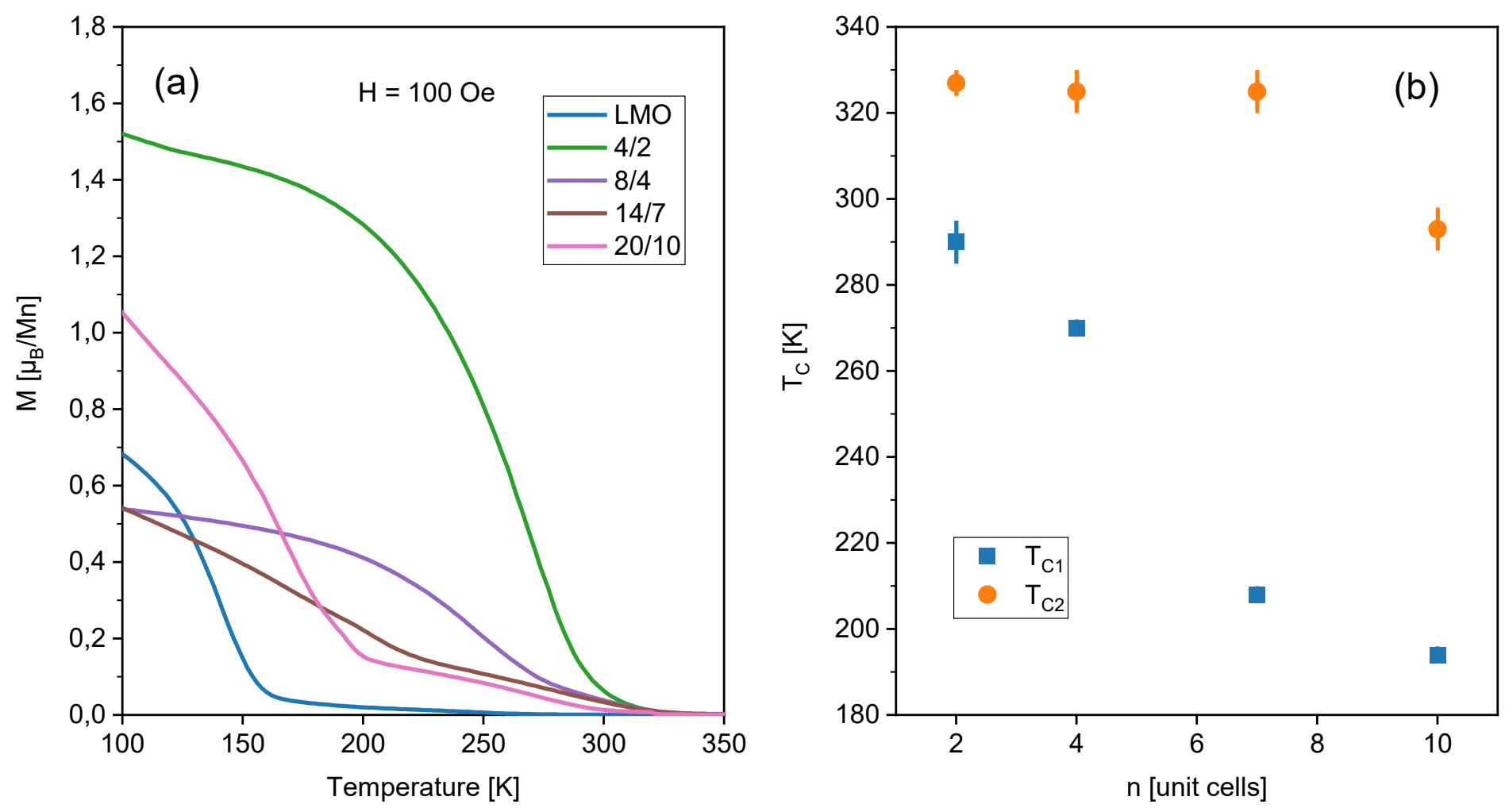

*Figure S8: (a) Temperature-dependent magnetization M(T) of $[LaMnO_3]_m/[SrMnO_3]_n$ superlattices measured under an external field of 100 Oe. (b) Curie temperatures $T_{C1}$ (bulk-like phase) and $T_{C2}$ (interfacial phase) as a function of $SrMnO_3$ layer thickness n.*

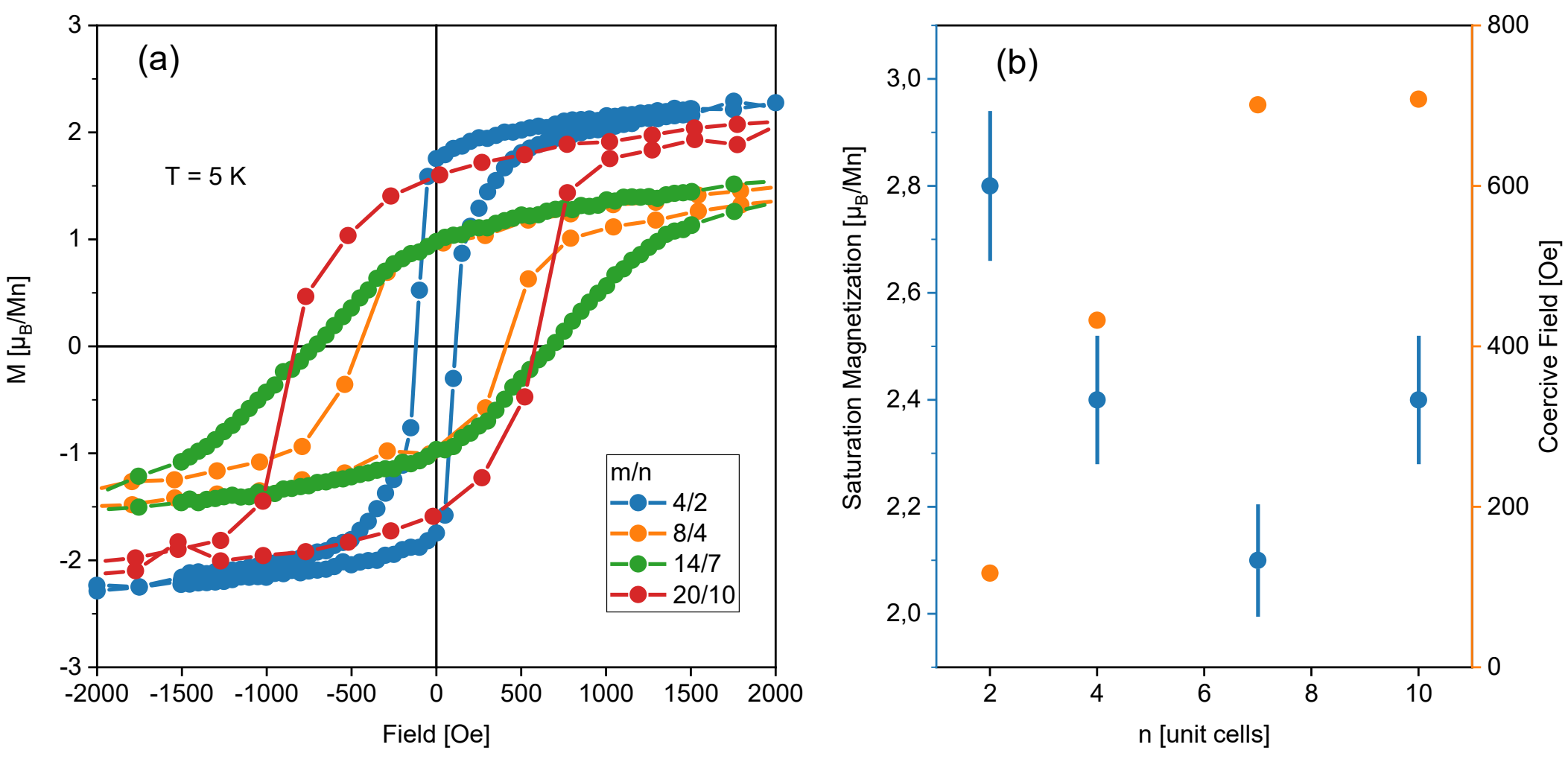

*Figure S9: (a) Magnetic hysteresis loops at $T = 5$ K for $[LaMnO_3]_m/[SrMnO_3]_n$ superlattices with m=2n. (b) Evolution of saturation magnetization (at $B_{ext} = 5$ T) and coercive field as a function of $SrMnO_3$ thickness n.*

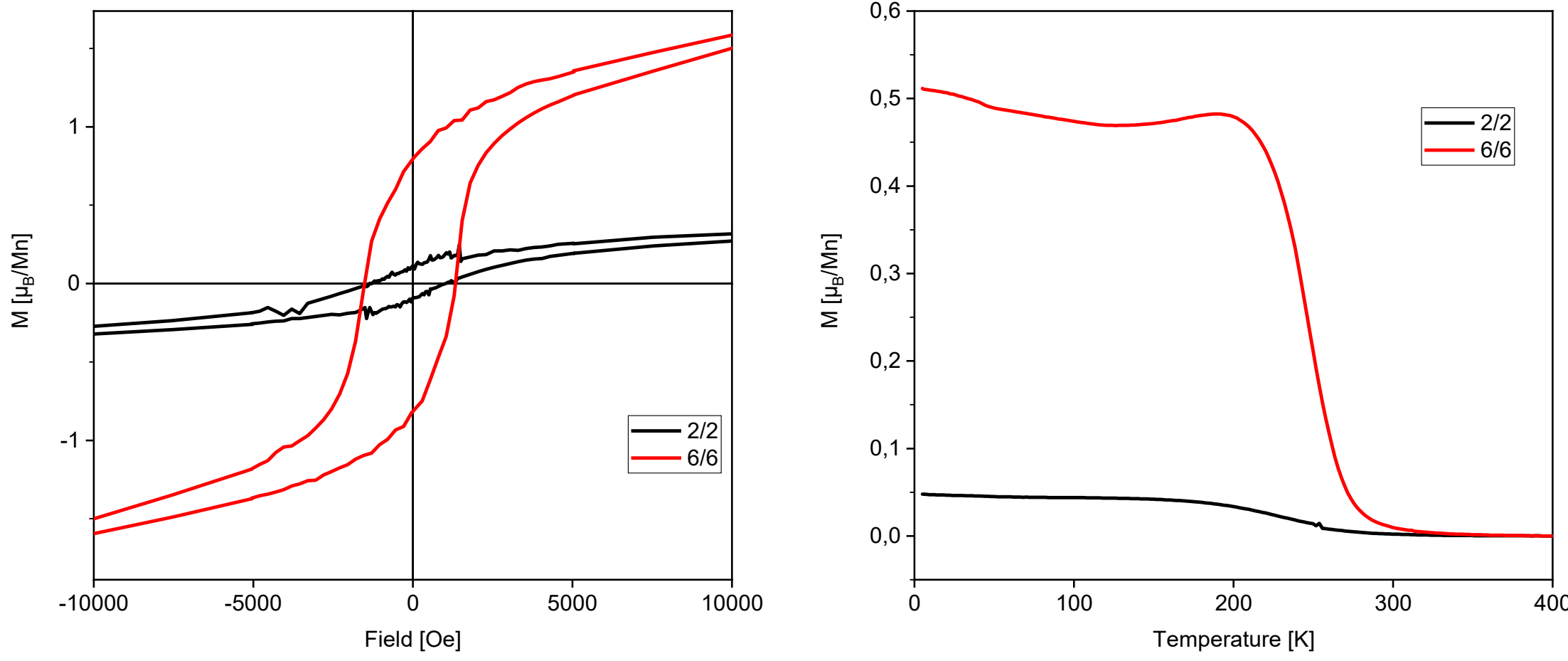

*Figure S10: Magnetic properties of $[LaMnO_3]_m/[SrMnO_3]_n$ superlattices for m=n: (left) hysteresis loops at $T = 5$ K; (right) field-cooled magnetization measured at $H = 100$ Oe.*

## SI-2 SEM Cantilever

To verify the dimensions of the cantilevers used in friction measurements, scanning electron microscopy (SEM) images (Figure S11) were acquired for a representative Nanosensors PPP-LFMR cantilever and tip.[10] The measured geometry and dimensions were found to be consistent with manufacturer specifications, confirming reliable input parameters for force calibration and contact area estimation.

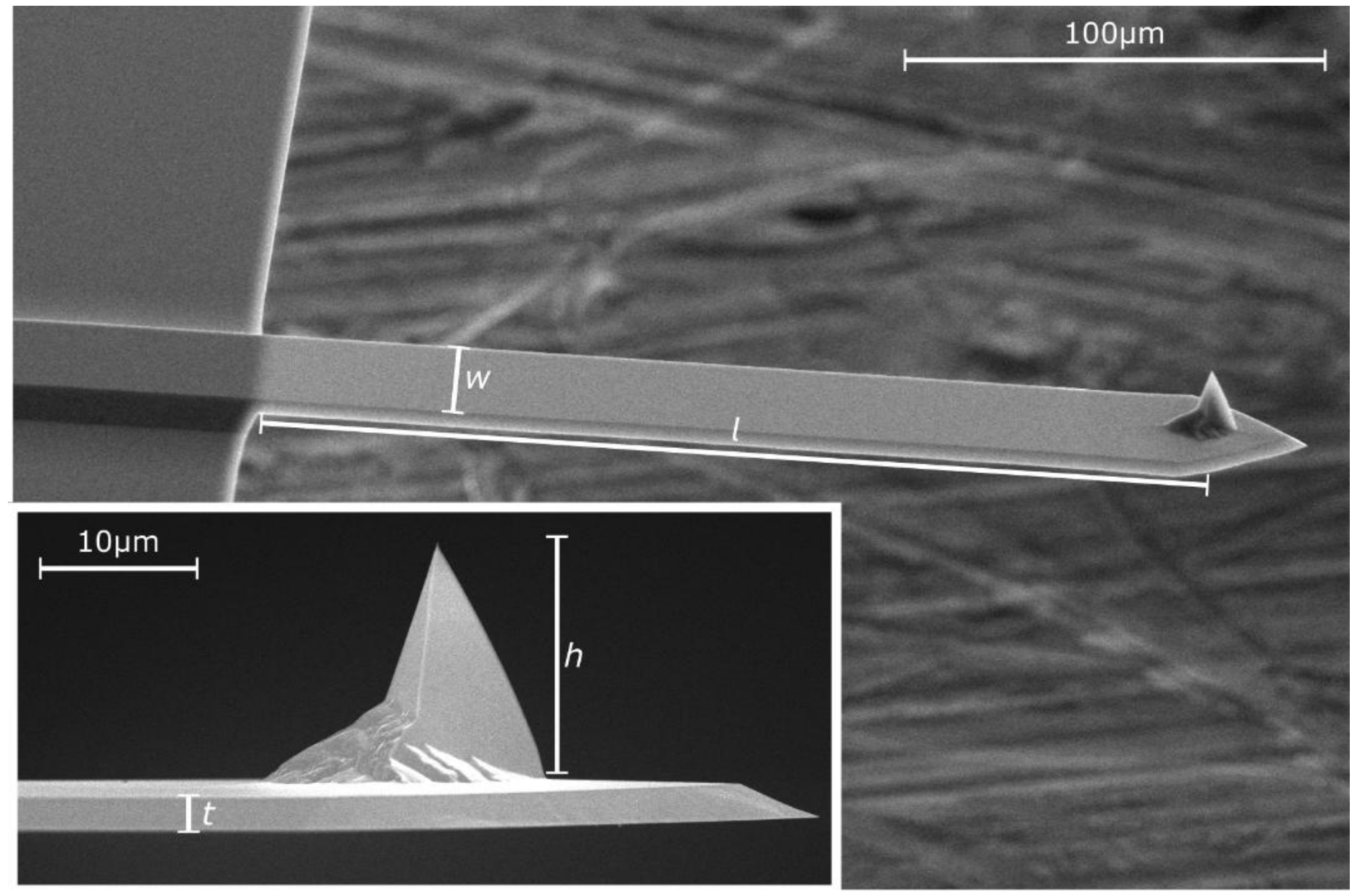

*Figure S11: Scanning electron microscopy images from a cantilever and tip (Nanosensors PPP-LFMR)[10] used to verify the dimensions stated by the manufacturer.*

## SI-3 Topography and Friction Maps

Atomic force microscopy (AFM) measurements were performed at room temperature (T = 293 K) under ultra-high vacuum (UHV, $p \approx 2\times10^{-10}$ mbar) using an Omicron VT-AFM/STM. For lateral force microscopy, standard rectangular single-crystalline silicon cantilevers (NANOSENSORS™ PPP-LFMR)[10] with a nominal tip radius R ≤ 10 nm were used. These cantilevers exhibit normal spring constants $k_n$ ranging from 0.88 to 1.3 N/m, and lateral spring constants $k_t$ ranging from 132.3 to 177.4 N/m. The lateral constants were calculated using geometric parameters provided by the manufacturer and literature values for material properties.[11] Normal spring constants were experimentally determined using the Sader method,[12,13] which relies on the cantilever's known geometry and its measured oscillation quality factor.

Friction forces ($F_F$) were extracted by averaging over 50 friction loops, each recorded within a 100 × 250 $nm^2$ region at a sampling density of 1 $nm^{-1}$ along the fast scan axis and 0.5 $nm^{-1}$ along the slow axis. To assess measurement reproducibility and spatial variability, friction experiments were repeated on randomly selected areas of each film. To minimize wear during scanning, the applied normal load ($F_N$) was kept below 30 nN, and possible changes in tip-sample contact were monitored through adhesion measurements.

To measure several films consecutively without the need to change the cantilever or the force calibration, up to four films were adhered to a Omicron stainless steel sample plate using a silver-filled epoxy (EPO-TEK H21D) to ensure good thermal and electric conductivity. Additionally, a small droplet of contact silver (ACHESON 1415 G3692) was placed at the edge of each film, far away from the area studied, to electrically contact the sample surfaces.

Representative topography and lateral friction maps from six $[LaMnO_3]_n/[SrMnO_3]_n$ superlattice films are shown below (Figure S12). All measurements were conducted at room temperature under UHV conditions, using a sliding speed of 250 nm/s and normal loads between 1 and 50 nN. All films exhibited an RMS roughness ≤ 0.5 nm with topographic features around 50 nm in size. In contrast, friction force maps revealed variations of up to a factor of four, with feature sizes ranging from 50 nm down to ~5 nm. The low correlation between friction and topography (Spearman correlation coefficient $r_s < 0.15$) suggests that friction contrast arises from intrinsic material variations rather than surface morphology. Some maps display sub-50 nm features in topography or horizontal trace lines in friction, which are attributed to tip wear or subtle changes in the contact geometry during scanning.

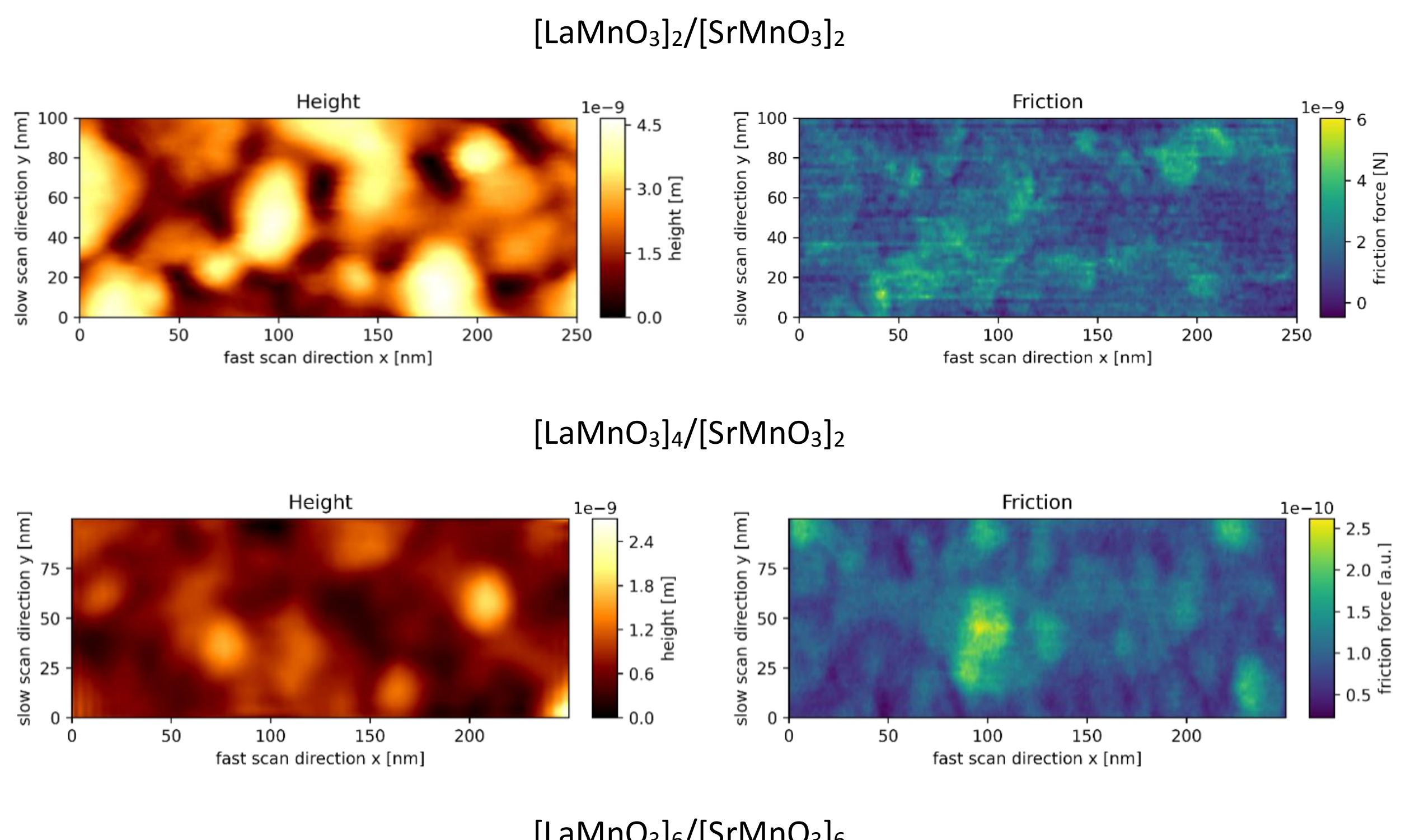

$[LaMnO_3]_6/[SrMnO_3]_6$

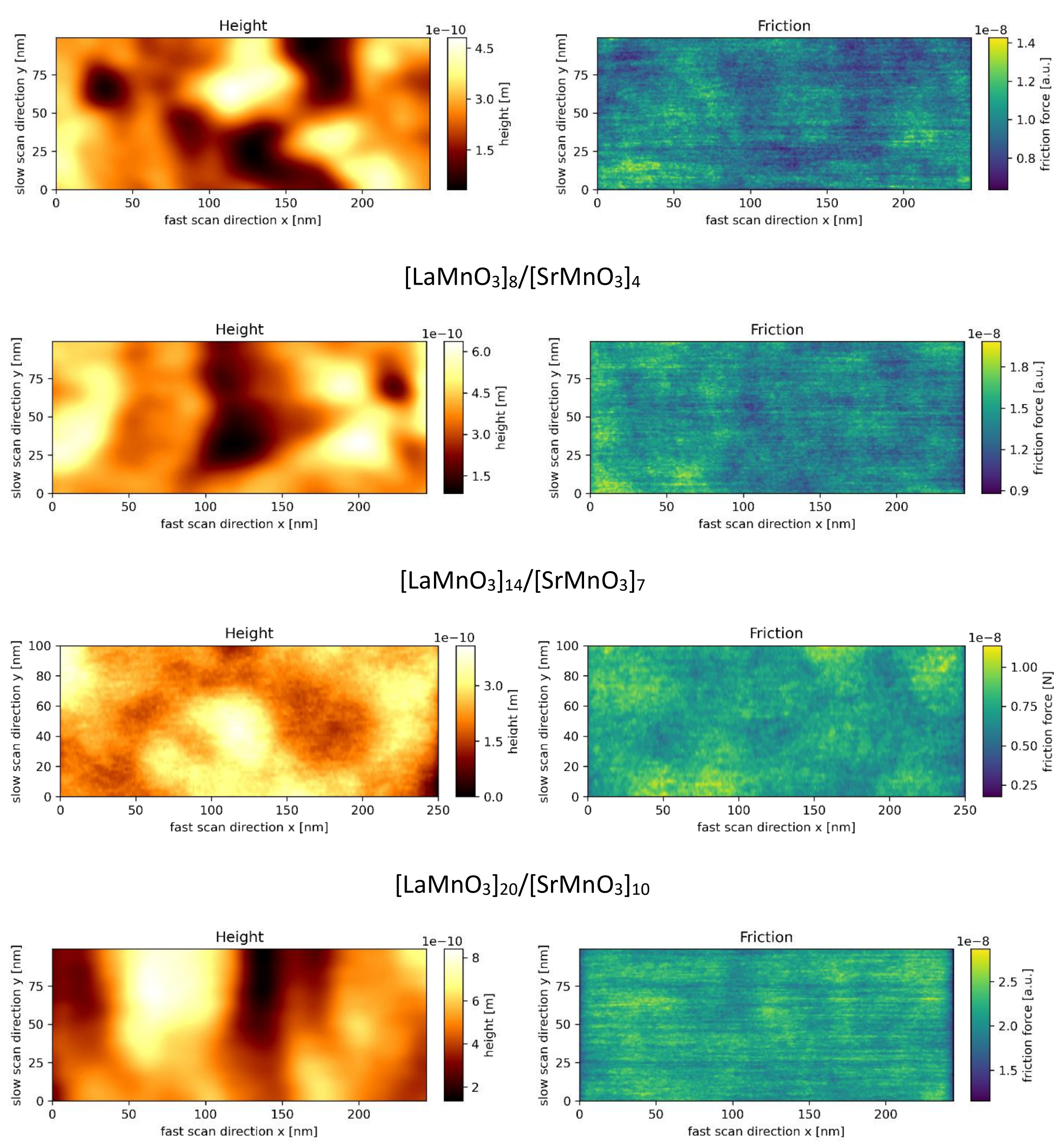

*Figure S12: (Left) Topography and (right) corresponding friction force ($F_x$) maps of the six $[LaMnO_3]_n/[SrMnO_3]_n$ superlattice films. All measurements were performed under UHV conditions using an Omicron VT-AFM at room temperature. There is no apparent correlation between friction contrast and topographic features.*

## SI-4 Friction versus Normal Force Measurements

### SI-4.1 Friction Force as a Function of Applied Normal Force

The plots below show the average friction force as a function of normal load for all six $[LaMnO_3]_m/[SrMnO_3]_n$ superlattice films. Each data point represents the average friction force extracted from a 100 nm × 250 nm friction map acquired at a specific normal load. A set of such measurements taken within approximately 1 µm of each other and across a range of normal loads constitutes one Amontons measurement set, which is represented by a consistent symbol style and fitted with a linear trend (dotted lines).

Each Amontons set was recorded at a randomly selected location on the film. Within a given set, the normal force was systematically increased and then decreased to the initial value to identify possible hysteresis or changes in tip condition, such as wear or contamination.

All Amontons measurements exhibit a linear relationship between friction force and applied load, consistent with Amontons' law when accounting for adhesive contributions:

$$F_F = \mu \cdot F_N + F_{F0}.$$

While the slopes of the linear fits (i.e., the friction coefficients $\mu$) are relatively consistent for each film, the intercepts $F_{F0}$ vary considerably. This indicates substantial differences in adhesion forces across different surface regions, reflecting local heterogeneity in tip-sample interactions. Notably, the three fluorine-contaminated samples (m = 2, 4, and 14) exhibit significantly lower friction coefficients. In contrast, the intercept values $F_{F0}$ show no clear correlation with film thickness or other structural parameters (see Figure S14).

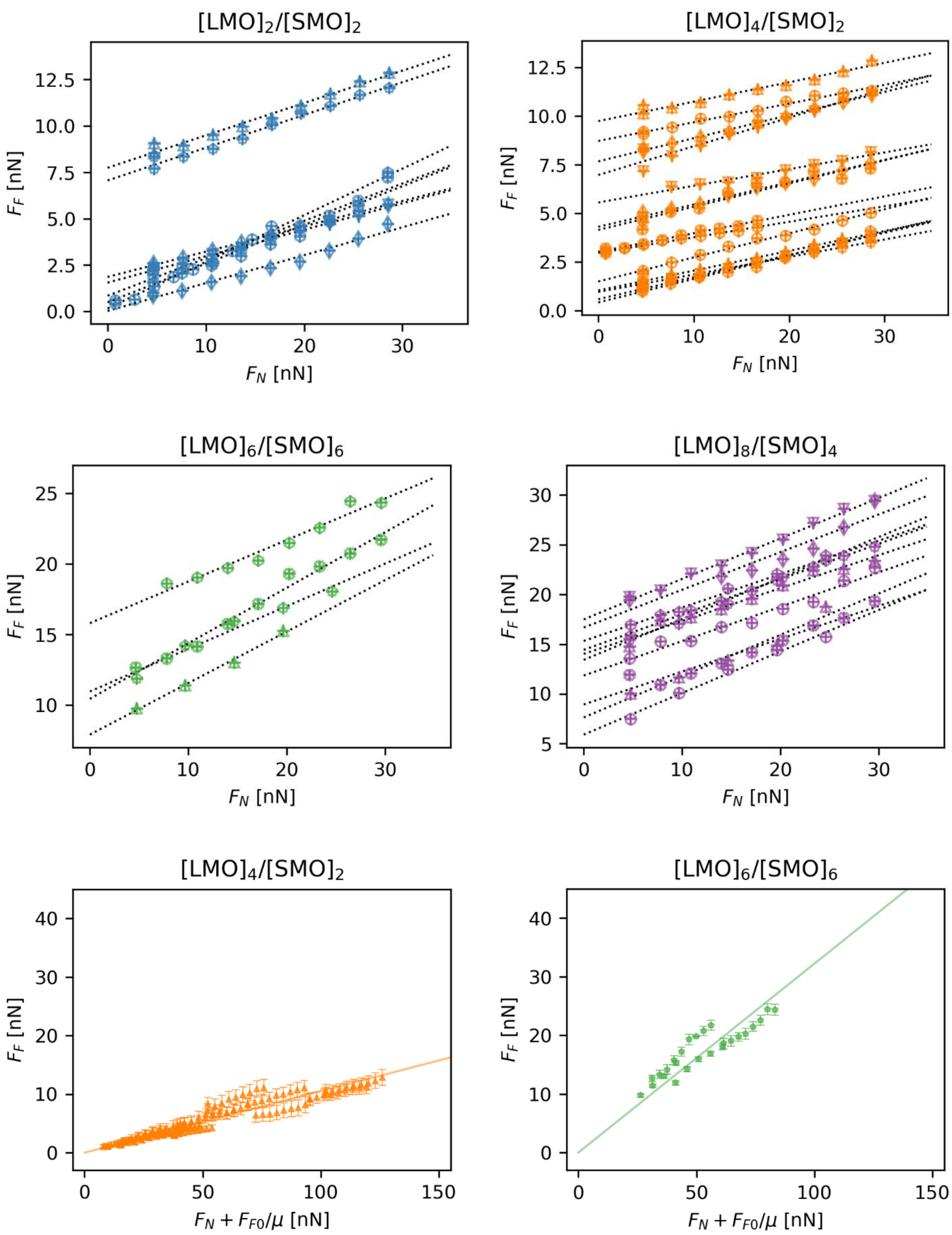

*Figure S13: Average friction force $F_F$ as a function of normal load for six $[LaMnO_3]_m/[SrMnO_3]_n$ superlattice films.*

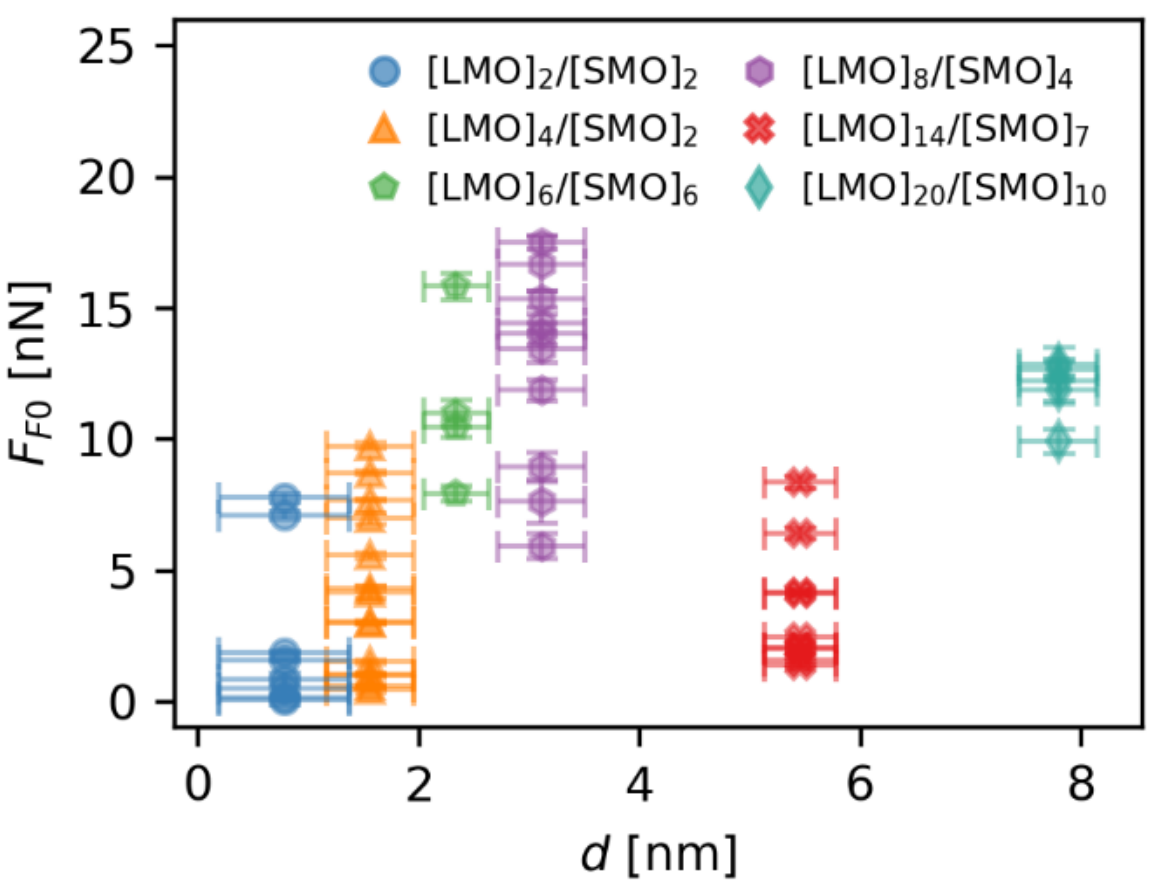

*Figure S14: Friction force intercept $F_{F0}$ versus top layer thickness for the six $[LaMnO_3]_m/[SrMnO_3]_n$ superlattice films. Each data point corresponds to an Amontons measurement performed on a localized region of the film surface. Films with fluorine contamination (m = 2, 4, 14) show consistently lower intercepts.*

### SI-4.2 Correlation Between $F_{F0}/\mu$ and Pull Off Forces

Contact mechanics models such as the DMT and JKR model[14,15] propose that friction vanishes when the applied normal force exactly compensates the adhesive interactions at the tip–sample interface. This implies that the ratio $F_{F0}/\mu$, which corresponds to the extrapolated normal force at which the friction force goes to zero, should be comparable in magnitude to the measured adhesion force. The latter can be estimated from the pull-off force obtained from force–distance curves.[16] The two forces are compared in Figure S15. Figure S15 compares these two quantities. The good agreement (dashed line indicates equality) confirms that the observed offsets in friction force $F_{F0}$ arise primarily from adhesive contributions.

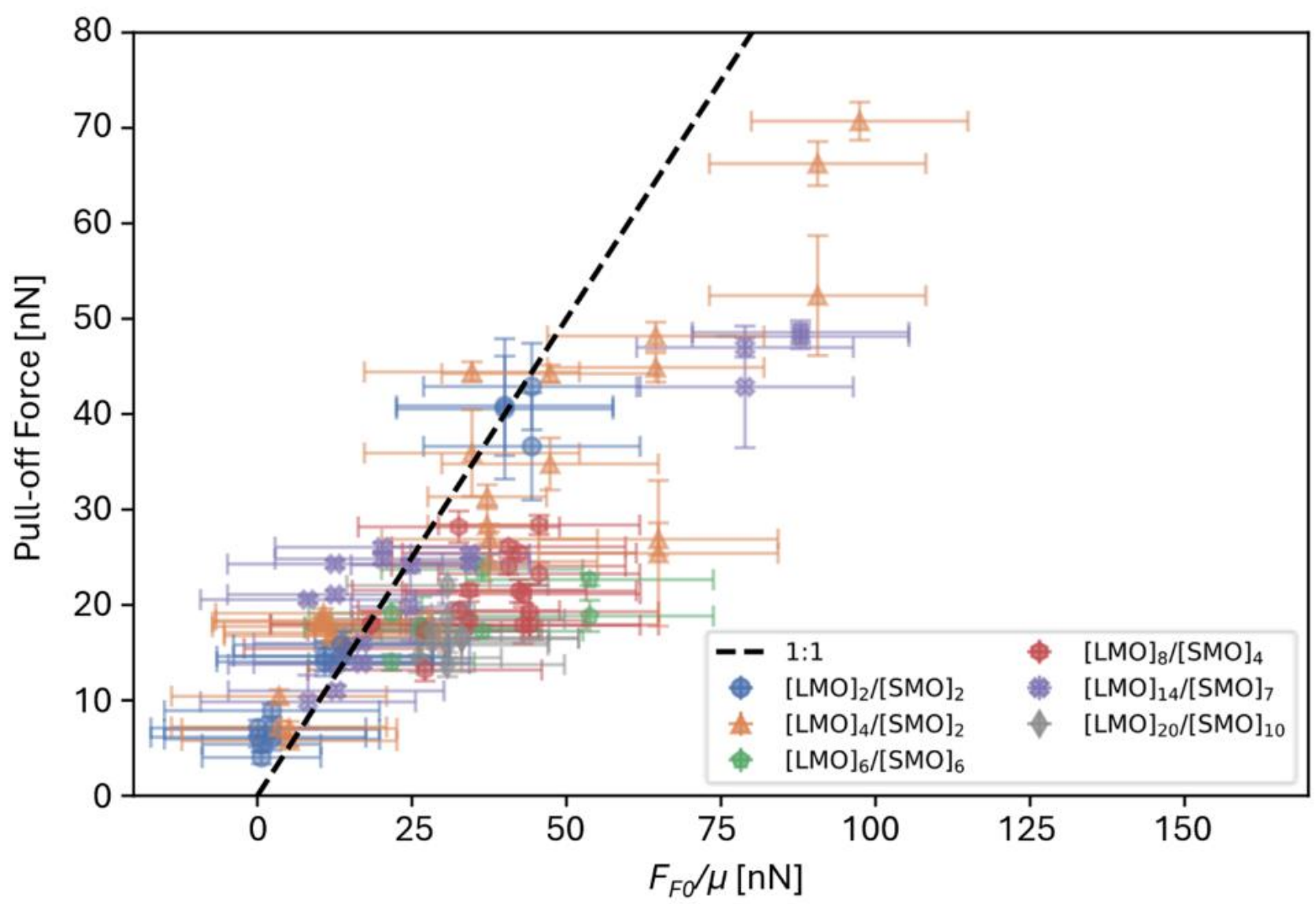

*Figure S15: Comparison of the adhesion force estimated from force–distance pull-off measurements and the extrapolated offset force $F_{F0}/\mu$ from Amontons plots. The dashed line represents perfect agreement. The close correlation supports the interpretation that adhesive interactions dominate the friction force offset.*

### SI-4.3 Friction Force as a Function of Applied Normal Force $(F_N + \mathrm{F_{F0}}/\mu)$

The close agreement between the pull-off force and the extrapolated normal force at zero friction supports the interpretation that adhesion is responsible for the vertical offset in the friction force, $F_{F0}$. When this adhesive force, $F_{F0}/\mu$, is added to the applied normal force $F_N$, the friction force data from Section SI-4.1 collapses onto a single linear master curve for each superlattice film (Figure S16). The slopes of these unified curves represent the average friction coefficients, denoted as $\overline{\mu}$.

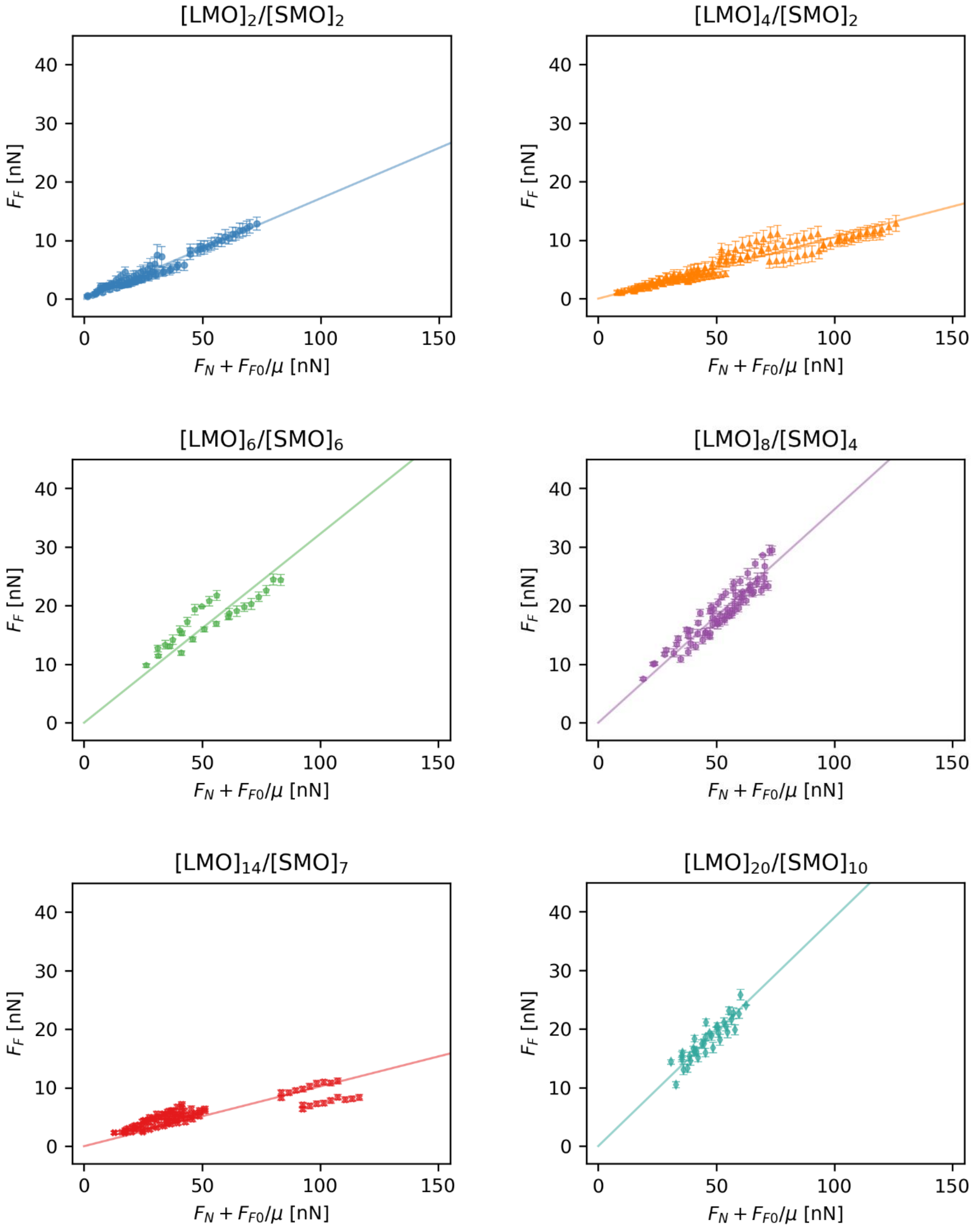

*Figure S16: Master curves for each superlattice film obtained by correcting the applied normal forces with the adhesive offset force $F_{F0}/\mu$. This correction collapses the data onto a single linear trend per film, with the slope defining the average friction coefficient $\overline{\mu}$.*

## SI-6 Bibliography